\documentclass[cernpreprint, atlasdraft=false, coverpage=false, UKenglish, backref=false, texmf, orcidlogo]{atlasdoc}
\usepackage{atlaspackage}
\usepackage{gensymb}
\usepackage{atlasbiblatex}

\usepackage{atlasphysics}

\usepackage{subfig}

\graphicspath{{logos/}{figures/}}

\AtlasTitle{The environmental impact, carbon emissions and sustainability of computing in the ATLAS experiment}

\AtlasAbstract{%
ATLAS, a general-purpose experiment at the Large Hadron Collider (LHC), makes use of a large internationally-distributed computing infrastructure, including over $10^6$~TB of managed data on disk and tape and almost one million simultaneously running CPU cores. Upgrades for the High-Luminosity LHC (HL-LHC) will increase the required computing resources by a factor of 3--4 by the beginning of the 2030s, and by an order of magnitude before the conclusion of data taking at the beginning of the 2040s. These resources are spread over around 100 computing sites worldwide. Efforts are underway within the experiment to evaluate and mitigate various aspects of the environmental impact of the sites, with the additional long-term goal of making recommendations to the sites that will significantly reduce the total expected environmental impact in the HL-LHC era. These efforts take several forms: building awareness in the experiment community, adjusting aspects of the computing policy, and modifications of data center configurations, either in ways that take advantage of particular features of ATLAS workloads or in generic ways that reduce the environmental impact of the computing resources. This paper describes the ongoing investigations and approaches that have already provided useful and actionable outcomes.
}

\AtlasRefCode{SOFT-2024-01}

\PreprintIdNumber{CERN-EP-2025-105}

\AtlasJournal{EPJC}
\AtlasJournalRef{\EPJC 85 (2025) 1397}
\AtlasDOI{10.1140/epjc/s10052-025-14976-3}

\hypersetup{pdftitle={ATLAS document},pdfauthor={The ATLAS Collaboration}}

\begin{document}

\maketitle

\section{Introduction}
\label{sec:intro}

As the world continues to experience the increasing effects of global warming~\cite{IPCC_AR6}, the sustainability, environmental impact, and carbon footprint of various aspects of our society have come under scrutiny. A series of efforts have been undertaken at CERN to characterize and reduce the energy, environmental, and carbon footprints~\cite{CERN-report-2020,CERN-report-2022} of the organization's activities. Computing is one component of the larger issue of sustainability in high-energy physics (HEP), and some aspects of this issue have been addressed by the Worldwide LHC Computing Grid (WLCG), the collaboration that encompasses most of the computing resources used by the LHC experiments today~\cite{WLCG-Energy}.

The ATLAS experiment~\cite{PERF-2007-01,GENR-2019-02} utilizes an extensive distributed computing system detailed in Ref.~\cite{SOFT-2022-02}; the planned expansion of these resources over the next 5~years in preparation for the High-Luminosity LHC (HL-LHC) upgrade is detailed in Ref.~\cite{CERN-LHCC-2022-005}.
In short, the experiment typically uses around 600~000--700~000 cores of compute, with peaks of more than one million cores.
Over $10^6$~TB of storage is in use, comprising over 400~PB of disk and 600~PB of tape.\footnote{The storage here is network-attached storage and does not include the scratch space generally associated with a compute node, which can be about 20~GB per core or an additional 20~PB worldwide. The compute includes active compute elements and does not include interactive compute, service nodes or compute associated with disk servers, amongst other items. Neither resource includes personal computing (e.g.~laptops).}
These resources are divided among about 100 distributed computing centers around the world, including WLCG sites, high-performance computing (HPC) centers, volunteer computing, and various other resources.

By 2030, when the HL-LHC upgrades are complete, the average required computing capacity of the experiment will be around 1.5--3 million cores; that number is expected to more than double by the end of the experiment's life in 2041.
Similarly, the disk storage needs are set to exceed 1--$2\times 10^6$~TB in 2031 and the tape needs to exceed $2\times 10^6$~TB, with both doubling again by 2041.
This growth of about an order of magnitude over 15 years implies a significant increase in the carbon footprint of the experiment's computing effort unless mitigated.
Given this long-term foresight of the implications of the HL-LHC upgrade for energy and carbon footprints, CERN and the WLCG have the unique opportunity to strategically mitigate the expected increase over several years, to grow computing resources in as sustainable a way as possible.

This paper summarizes the current understanding of the likely evolution of the carbon footprint of future compute demand
and the ongoing efforts to reduce the overall environmental impact of computing for the ATLAS experiment, through both policy and direct action.
Some of these efforts can be directly translated into recommendations for site administrators, developers, and users towards this reduction.
These studies are intended to help inform choices in the computing model to be used in the HL-LHC and to guide sites in their purchase decisions.
The recommendations and policies are also intended to reduce the environmental impact of computing without having any meaningful impact on the physics program of the experiment.
The policies, recommendations, studies, and examples may also prove useful to other experiments considering ways to reduce the environmental impact of their computing.

The sustainability efforts within ATLAS computing can be divided into four groups.
First, as described in Section~\ref{sec:awareness}, efforts have been made to improve awareness among users, developers, site administrators, and managers of issues around sustainability.
Second, experiment policies and standard practices can be adjusted to minimize the environmental impact of computing, often without any harm to the physics program of the experiment.
These policies are discussed in Section~\ref{sec:policy}.
Third, there are some actions that site administrators can take to take advantage of the particular patterns and features that ATLAS workloads and site use follow.
These actions are described in Section~\ref{sec:atlasactions}.
Fourth, there are several actions that site administrators can take themselves to lessen the environmental impact of their site.
This last category includes things that are not specific to ATLAS, or even to high-energy physics, but are of general applicability.
These ideas are being studied and developed by many other groups, like the EE HPC WG~\cite{EEHPCWG}, the Open Compute Project~\cite{OCP}, Green DiSC~\cite{GreenDisc}, GreenDIGIT~\cite{GreenDIGIT}, the UKRI Net Zero Digital Research Infrastructure Scoping Project~\cite{UKRI}, Boavizta~\cite{Boavizta}, or the International Energy Agency~\cite{AIEnergy}, and several ATLAS sites have been deeply involved as well.
The goal is therefore not to replace or reproduce, but to complement the existing efforts outside of ATLAS.
The ongoing investigations associated with ATLAS sites are described in Section~\ref{sec:sites}, and possible improvements through knowledge sharing are described in Section~\ref{sec:sitesharing}.
All of these efforts and approaches must be pursued if the environmental impact of the collaboration is to be minimized.

The most common mechanism for estimating organizational carbon emissions is the Greenhouse Gas (GHG) Protocol Corporate reporting standard~\cite{GHG}.
This protocol divides carbon emissions into three ``scopes'':

\begin{enumerate}
\item Direct emissions, including fuel consumption, refrigerants, chemicals, and gasses;
\item Mainly indirect emissions due to generation of electricity used on site, and also due to the acquisition of steam, cooling or heating from offsite; and
\item Other indirect emissions, a catch-all for any other emissions that occur in an organisation's value chain, including up-stream and down-stream activities.
\end{enumerate}

Although computing centers have minimal direct (Scope 1) emissions, chip manufacturers' Scope 1 emissions may be considerably larger. The Scope 1 and 2 emissions of computing hardware manufacturers are generally accounted for as a part of the Scope 3 emissions of a data center. The cost associated with the production of hardware is referred to as ``embodied'' or ``embedded'' carbon.

Scope 2 electricity emissions can be reported as ``Location-based,'' depending on the mean carbon intensity of the grid in the location where it is used, or ``Market-based,'' reflecting emissions based on intensities of contractual purchases of electricity. Here location-based emissions are used, because they more directly reflect emissions associated with electricity use. Scope 2 emissions present an interesting challenge for ATLAS computing, because the resources are distributed over the world, and the large number and variety of sites each have different power efficiencies, energy practices, and power grid properties. These all vary with time, as well, and the decarbonization of the power grid in various countries is important to account for in extrapolations of the carbon footprint of computing. The lifetime model of typical WLCG sites is also important to consider: sites tend to retain and operate hardware well beyond its period of warranty, producing a significantly different amortization for embodied carbon compared to a commercial data center.

\begin{figure}[htb]
\centering
\includegraphics[width=0.95\textwidth]{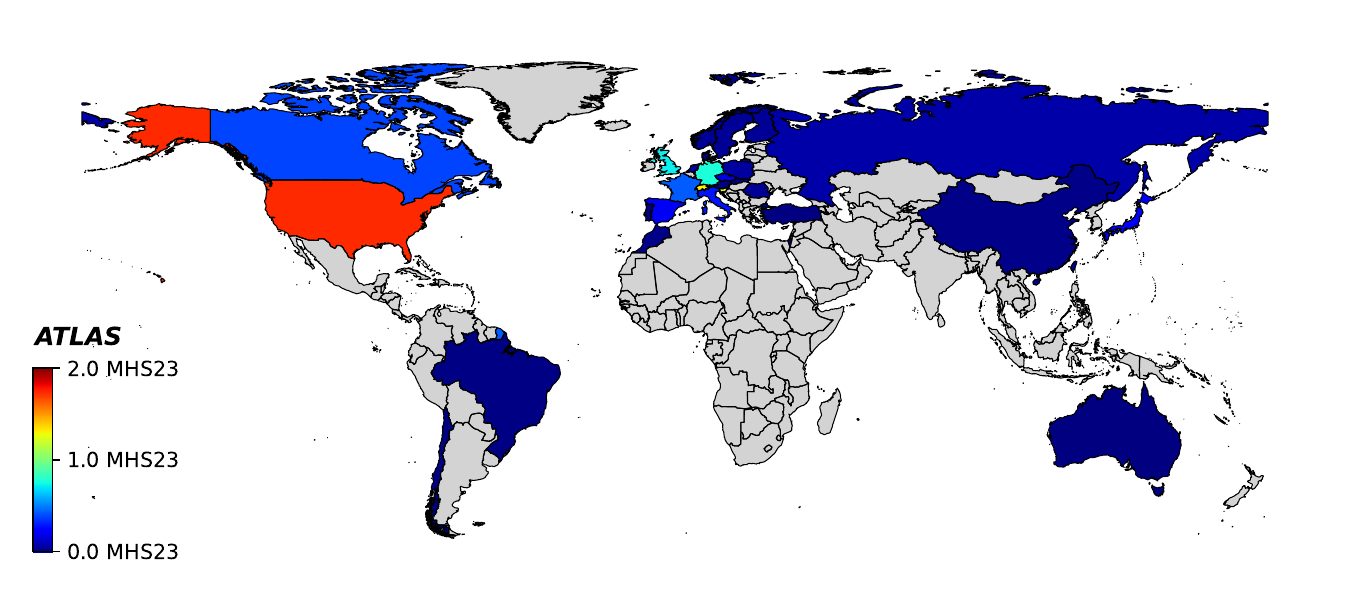}
\caption{The worldwide distribution of ATLAS computing, based on the amount of CPU provided, in HS23 (see Section~\ref{sec:HS23} for the precise definition of HS23), on average in 2023--2024. Countries in gray did not contribute significant CPU.}
\label{fig:sitemap}
\end{figure}

CERN has estimated the Scope 2 emissions from its own computing facilities at around 5\% of the total footprint of the laboratory when the LHC is operating, with the majority coming from the cooling of the accelerator complex~\cite{CERN-report-2020,CERN-report-2022}. As CERN operates about 15\% of the ATLAS computing resources, and many non-CERN resources rely on power grids with significantly higher carbon intensity than that of France (from which CERN gets its power), the total Scope 2 carbon footprint of the ATLAS computing resources may rival that of CERN itself. The worldwide distribution of ATLAS computing is shown in Figure~\ref{fig:sitemap}, based on the amount of CPU provided in HS23, the computational power based on the 2023 HEPScore benchmark (see Section~\ref{sec:HS23} for the precise definition of HS23).

One critical metric in the discussion of power consumed by computing is the Power Usage Effectiveness (PUE), which is defined as the ratio of power supplied to the building housing the computing resources to that used for computing (i.e.~useful energy). Typical PUEs for sites on the WLCG are about 1.5~\cite{WLCG-Energy}; a ``good'' PUE is 1.1, which has been achieved at the new CERN data center, for example~\cite{CERN-report-2022}. For most of this paper, the units used for estimation are kilograms of CO$_2$ equivalent, or kgCO$_2$eq. This is the standard unit used for carbon footprints, and requires the translation of all gas outputs into a single unit.

The most commonly discussed and estimated aspect of the environmental impact of computing is due to the power consumption of the hardware and corresponding cooling.
Historically, operational (location-based Scope 2) emissions of large data centers have been reported as larger than Scope 3, which originate in particular from hardware manufacturing and infrastructure.
More recent studies, however, indicate that the latter may be more significant, especially as grid carbon intensities reduce and more comprehensive and accurate embodied emissions methodologies and data become available (see, for example, Ref.~\cite{2011.02839}).
The implication is that other aspects of the carbon footprint and environmental impact of computing, beyond the power consumption, cannot be overlooked when building a more holistic, forward-looking and comprehensive model. Recommendations based solely on the impact of changes in power consumption, which ignore other aspects like Scope 3 emissions or power grid decarbonization, may prove misleading and inaccurate when a more complete model is considered.\footnote{In principle, this discussion applies equally to the estimate of the environmental impact of other aspects of ATLAS. However, the focus here is on the relative components of the environmental impact of computing, rather than in the impact of computing relative to the impact from other contributions.} Estimates of embodied carbon often come with significant uncertainties, due to deficiencies or difficulties in manufacturer reporting. As this work progresses, even a comprehensive corporate emissions assessment of Scopes 1--3 might not be sufficient when evaluating the global environmental impact of computing, which also includes water consumption, the use of large volumes of rare or difficult to extract elements, waste management and other socio-environmental impacts~\cite{2011.02839}. In addition, utilising carbon impact assessment methods more suited to informing policy development and design changes may become more appropriate than the standard corporate reporting approach. For this paper, however, the focus is left on the corporate carbon footprinting approaches, in particular including Scope 2 and Scope 3 emissions associated with embodied emissions of equipment and data center construction.


%
\section{Improving awareness}
\label{sec:awareness}

One of the most important ways to reduce waste in the environmental impact of ATLAS computing is to improve the awareness of environmental issues, and to make members conscious of the positive impacts their decisions and actions can have.
Workloads on the grid are classified as either ``production'' jobs, which include data processing and Monte Carlo simulation dataset production, are run by experienced production teams, and usually follow well-validated patterns, and ``user'' jobs, which are submitted by any users in the collaboration and may contain almost any code.
User jobs typically comprise 10\% of ATLAS computing HS23 and 30\% of jobs; the jobs are much shorter on average than production jobs.
Problematic user workloads are a significant source of waste, and the efforts towards education are meant to serve as a reminder of the consequences of choices made --- not to shame users, but to remind them that their decisions are important.

To that end, PanDA~\cite{PanDA}, the workflow management used by ATLAS, has been equipped to provide estimates of the carbon footprint of all ATLAS jobs.
The system draws power grid carbon intensity in kgCO$_2$eq/kWh for each site from Electricity Maps~\cite{ElectricityMaps}, and sites can provide an estimate of the number of Watts per core of compute at their site.
This estimate is a stand-in for the entire power draw of a job: the compute itself, as well as the associated share of cooling, storage, networking, and any other relevant hardware.
A default value of 10~W per core was assigned, based on Ref.~\cite{GreenAlgorithms}.

Users are then provided an estimate of the carbon footprint of their job, along with a link to more information about the calculations~\cite{PanDACarbon}.
An example of the report that a user receives is shown in Figure~\ref{fig:carbon_email}.
Rather than using the site-specific power grid carbon intensity, the reported carbon footprint estimate uses an average of the global power grid carbon intensity across all ATLAS sites.
This is for several reasons:

\begin{figure}[ht!]
\centering
\includegraphics[scale=0.7]{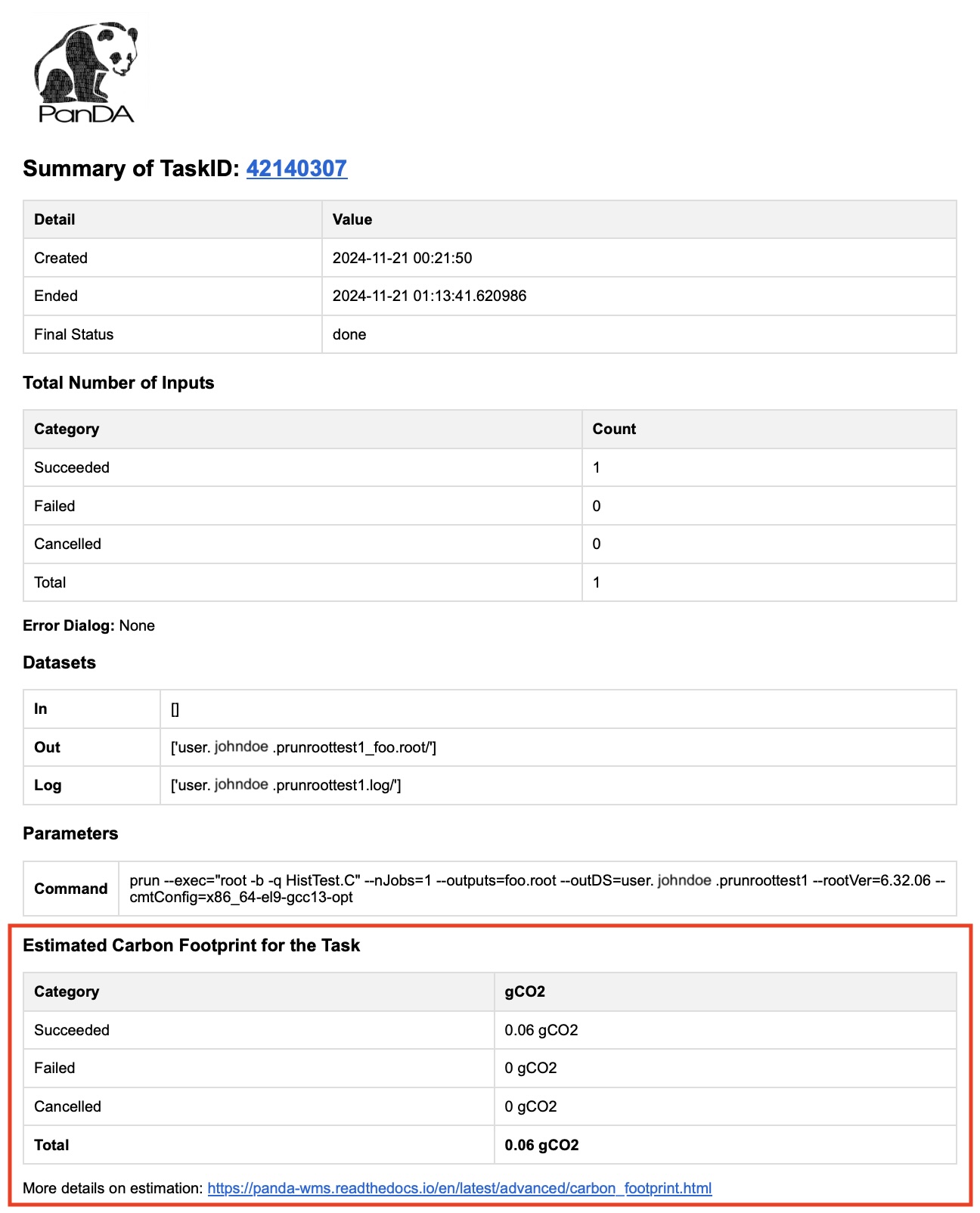}
\caption{An example email of a job report from PanDA including carbon footprint information. The carbon footprint information is boxed at the bottom of the example email, and includes a link to a website providing more information about the calculations~\cite{PanDACarbon}.}
\label{fig:carbon_email}
\end{figure}

\begin{enumerate}
\item Site data is still in its infancy and is relatively unreliable.
For example, sites may have solar panels installed or may be paying a premium for renewable energy.
The PUE of sites varies significantly, and it is likely that 10~W is an overestimate for some sites and an underestimate for others.
Until the site data are complete, the average avoids assigning an incorrect impact to a site.
This 10~W estimate also stands in for the power consumption of other IT elements required by the jobs, including network and storage.
A more detailed understanding of the real power consumption is required before this can be considered truly reliable.
\item Using the average ensures a more direct connection between the optimization of a user's code and the carbon footprint of that code.
Faster running code results in a smaller carbon footprint, ignoring global time variations of power grid carbon intensity.
This is important to correctly incentivize and reward positive user actions.
\item Users might respond to site-specific data by redirecting their jobs to sites with lower power grid carbon intensity.
In fact, computing resources are not left idle: any cores not taken by user jobs will be consumed by production jobs (e.g.~Monte Carlo simulation).
Thus, users redirecting their jobs will not affect the number of cores of computing running in any region.
Instead, in fact, the impact might be negative: many users redirecting jobs to a single site will cause significant network usage, churning of data on disks, and congestion at the site.
The result may be both a longer waiting time for the users and a larger overall environmental impact for the experiment.
\end{enumerate}

\noindent Nevertheless, the site-specific numbers are recorded, and are available for internal study and improvement.
Although the resources will not sit idle (i.e.~if the user job is faster, other workloads will be run), this gives users a useful metric for the carbon footprint of their scientific result.

To ensure that new users are accustomed to seeing these reports, the standard ATLAS analysis software tutorial that most new members attend includes submission of a failing job.
The resource waste is minimal, but ensuring that users have some experience seeing what a failing job looks like, analyzing it and understanding the source of the failure is critical.
There were cases identified of users retrying failing jobs an enormous number of times, and the educational program is intended to help avoid such situations in the future.
To avoid substantial waste from large numbers of retried jobs, several cutoffs have also been introduced.
Configurable limits have been set on the allowed fraction of failed jobs and HS23s, as well as maximum numbers of task retries and attempts for an individual job.
Additionally, tasks with very low CPU efficiency are stopped, to prevent jobs from reserving more CPU cores than they require.

The same feature in PanDA can be used to provide reports for large production campaigns run by the experiment.
Production managers can see the reported carbon footprint for their preferred selection of jobs.
These reports separate the carbon footprint of successful and failing jobs: when a campaign has a significant failure rate, the environmental impact of the failures is visible.
Similarly, if a campaign has to be abandoned or redone because of a major software issue, the environmental impact of the lost production is visible.
These reports are not meant to prevent the generation of useful data for the experiment, but rather to ensure that appropriate care is taken in validating and setting up new production campaigns.

In a similar manner, the recasting of many experiment activities in an environmental sustainability light has been beneficial towards ensuring that more experiment members consider them thoroughly.
Software validation, continuous integration testing, and the validation of the physics output of new software releases can all be understood in sustainability terms: issues missed in validation are costly.
Historically, there were cases of software bugs identified far too late into production, only after many millions of events had been processed.
The cost of appropriately scaled validation (e.g.~frequently-run fast jobs, infrequently-run slower jobs) is insignificant in comparison to the CPU time lost to production with bugs.
Furthermore, attempts to optimize software are useful for resource savings, but are also relevant in environmental terms.

Towards raising awareness in site administrators, a survey was done to understand current practices across both large and small sites.
One of the issues ATLAS faces is that while most resources are provided by a relatively small number of large sites, the site count is dominated by many small sites, including small computing clusters at institutes.
These small sites often lack the capabilities of large sites: more than half of the computing sites surveyed do not monitor their electricity usage, while all the major WLCG sites do.
More than 60\% of the site administrators indicated a desire for advice and the sharing of best practices towards reducing environmental impact.


%
\section{Experiment computing policies}
\label{sec:policy}

The ATLAS computing model has many parameters and policies that can be adjusted that affect the environmental impact of the experiment.
To date, these policies and the model itself have been mostly examined through the lens of financial savings.
Often financial and environmental benefits are connected: simply reducing overall resources is beneficial in both respects.
Changes to the policies of the experiment are often the most effective towards sustainability: they can be taken immediately, without requiring site changes or new purchases, for example.

Policies often allow the experiment to consume one resource type and save another, sometimes with an extremely beneficial trade-off.
One simple example is the compression of data.
Compressing ATLAS data is an obvious thing to do: the enormous savings in storage, often a factor of three or more, hugely outweigh the modest CPU required to compress and decompress the data (1\%--10\% of the run time, depending on the workload).
However, once the initial step has been taken, there is still considerable room for further optimization: the compression algorithm can be changed, with the constraint that it must be sufficiently robust to not risk the long-term integrity of the data, and the algorithm settings can be changed to find the best working point given the expected frequency with which the data will be read~\cite{Compression}.
The data layout in storage can also significantly affect the resources required by the experiment.
Lossy compression, the dropping of insignificant bits, can save 5\%--15\% of the storage space required by data analysis formats, but requires careful validation to ensure that physics outputs are not distorted.
The adoption of the new ROOT RNTuple format, which has no loss but only improves the on-disk data layout, has the potential to save the experiment 15\%--20\% of its storage space, and potentially even more for frequently-read analysis data formats~\cite{RNTuple}.
These changes and their rapid adoption after validation can have a significant impact on the experiment's computing resource needs and thereby on its environmental footprint.

One example of a less obvious connection between financial and environmental benefit is the ``data carousel''~\cite{SOFT-2022-02} used by the experiment.
The data carousel is a means to using tape storage, which is normally purely archival, as a more active storage medium.
Infrequently used datasets are archived to tape, deleted from disk, and recalled in the future if they are required.
This reduces the overall disk requirement of the experiment at the cost of an increased tape storage requirement.
Because tape is a less carbon intensive storage medium per TB in terms of both operational carbon and embodied carbon~\cite{TapeFootprint}, the policy also reduces the overall environmental impact of the experiment.

Generally speaking, the resources in a single year suffer from the Jevons paradox~\cite{JevonsParadox}: the available compute is intentionally kept busy with scientific work that offers useful improvements of results.
This is not only due to the increased demand that can be satisfied by faster or more efficient workloads, but also to honor the cost of investment of the funding agencies.
Racks of CPU are expensive, and it is generally thought wise to always keep them busy with lower priority work that provides at least marginal scientific benefit rather than allowing them to sit idle.
Idle CPU presents an opportunity to improve the scientific output of the experiment.

Over longer timescales, however, resource requests are made by the experiment based on extrapolations of resource needs.
If some software is made significantly more efficient or faster, the resources will still be fully utilized, but a more modest expansion will be requested for subsequent years.
It is thereby possible for software optimizations to reduce the total carbon footprint of the experiment over the long term.
Continuous investment in software optimization and efficiency improvements is therefore valuable for environmental, as well as financial and scientific, reasons.
These efforts continue to pay dividends: the efficiency of 8-core Monte Carlo simulation production jobs, measured by the ratio of CPU time to wall-clock time, has risen by about 5\% over the last three years, to over 90\%.
The time per event required by Monte Carlo simulation has also been reduced by almost 50\%, thanks to the validation and deployment of a wide variety of optimizations~\cite{SOFT-2022-02}, which is reflected in reduced need for CPU at the WLCG sites.
Similarly, significant improvements in event generation software have led to substantial improvements, in some cases 50-fold speed improvements~\cite{Bothmann:2022thx,Bothmann:2023ozs}.
With such substantial speed improvements, the collaboration can decide how much to reinvest in samples with greater physics precision, and by how much to reduce the overall computing resource budget.

\subsection{Data reproduction}

Another example of an experiment policy that has both a financial and an environmental impact is related to data reproduction, particularly for data formats produced for the final stages of data analysis, called derived analysis object data, or DAODs.
These data formats are individually small, but often together comprise more than 25\% of the total data that ATLAS retains on disk.
New versions are frequently produced, in some cases several times per year due to software or calibration improvements, and old versions are deleted when they are no longer needed.
When data analysis teams are close to completing a publication, they will often ask that the particular version used for the preparation of the publication is retained.
Often this retention is ``just in case'': if during collaboration or journal review the analysis team is asked to check an issue that requires accessing DAODs, they need access to these files.
Because these files will be deleted in some relatively short period, it is ineffective to migrate them to tape: tape repacking is a slow process, and frequent deletion of small datasets from tape creates operational problems at sites.

To reduce the disk footprint of these files that are retained ``just in case'', functionality was built into the distributed production system that allows the reproduction of a dataset on demand.
This functionality can deliver an identical copy of the deleted data, in a time not much longer than what would be required to recall the data from tape.
This 1--2 day production, for modest dataset of a few TB, is sufficiently fast to not induce significant delays in the publication review.
The question remains, however, whether the use of dataset reproduction is beneficial from an environmental standpoint.

Because of their different embodied and operational carbon footprints, the relative cost of disk and compute varies by site.
Based on data from several sites (see also Ref.~\cite{Boavizta}), the number of Watt-seconds per CPU core-second is about 100 times the number of Watt-seconds per kB-year of storage.
Typical compute resources require about 7.3~W~/~core,\footnote{This differs somewhat from the 10~W/core discussed in Section~\ref{sec:awareness} because the 10~W/core estimate is a stand-in for additional computing elements like network and job-associated storage, and because this uses site-specific power estimates.} and typical storage requires about 0.06~Ws~/~kB~y (see for example Ref.~\cite{Spiteri}).
That means that, based on the average size of the data formats (30~kB/event) and time for production (20~HS23 s/event including all overheads, or about 1.6~s/event on a typical processor; see Section~\ref{sec:HS23} for more details on HS23), the creation of an event takes about 11.7~Ws of energy, compared to 1.9~Ws of energy per year to store the event.
The break-even point of a replication strategy is the point at which the compute and storage have equal cost.
In this case, the break-even point is a replication rate of $1.9/11.7=16\%$ for a year of prolonged storage.
In other words, saving the data for an additional year costs the same as reproducing 16\% of the data.
Therefore, any reproduction rate below 16\% would mean that reproduction on demand is an environmentally effective solution.
If the data are accessed many times, would need only to be saved for a short period, or would have to be reproduced with high probability, reproduction on demand is environmentally harmful.

\subsection{Automatic computational waste reduction}

The experiment has several policies and automated systems that help to automatically reduce computational waste (unproductive or invalid compute jobs).
This can be achieved in several ways, for example by taking sites offline if they are problematic, or by stopping tasks (collections of jobs with the same configuration and either different random number seeds or different input files, for example) if they are problematic.
One example of the first is HammerCloud~\cite{HammerCloud}.
Among its functions, HammerCloud is used to continuously test ATLAS computing sites with a variety of well-understood jobs that are expected to succeed.
About 0.05\% of the total CPU used by ATLAS in 2023 was used for these functionality tests.
If HammerCloud detects a problematic number of these test jobs failing, it automatically takes the site offline and notifies the administrators in a process called ``auto-exclusion''.
Were the site to remain online, large numbers of jobs directed to it would fail.
If the compute at the site sits idle, the power consumption of the CPU is reduced by about 50\%.
If the compute can be used for other purposes (e.g.~if a failure only affects ATLAS production jobs, and local user batch system jobs can still be run), then the entire power consumed by the CPU can be considered saved (i.e.~not wasted).
HammerCloud takes sites offline for about 4.6\% of a year on average through auto-exclusion, which can be thought of as an upper bound on the reduction in total energy required thanks to its application.
Several issues can reduce its effectiveness, like bad nodes that cause an otherwise good site to be taken offline, or the inability to fill the CPU with other useful work while it is offline for ATLAS.
The key to this savings is accurate, precise, and timely auto-exclusion, which is the focus of the ongoing sustainability efforts within HammerCloud.

Within PanDA, there are additional ``scout'' and ``retry'' schemes that are designed to prevent waste by reducing failures within a task.
For each task, the first 10 jobs, normally a small fraction of the total, are considered ``scouts'' and are run before the remaining jobs are released into the system.
If these scout jobs fail at a high rate, the task can be stopped before significant resources are consumed by failures.
The scouts can also be used to estimate the resource needs of the jobs in terms of CPU, memory, and disk space, so that jobs can be more optimally packed into existing computing resources.

When jobs fail, the failures are diagnosed and compared to a set of rules for retrying jobs.
Automatic retries help enormously in reducing the load on production managers that would otherwise have to take manual actions.
The rules ensure that if a job fails for a reason that appears to be site-related, like a corrupted input file, it can be moved to a different site; if the job fails for a reason that appears random, like an issue that could have been due to memory pressure from other jobs on the node, it might be retried 3--4 times; and if the job fails for a reason that is clearly reproducible, the job is not retried further unless experts intervene.
The rules for retrying jobs need regular review and maintenance, but they can help reduce the amount of waste in the system when properly tuned.

\subsection{Unused data}

Another effort designed to reduce waste is an examination of unused data.
These are data that, to the knowledge of the distributed data management system, have never been used as input for a job, downloaded by a user, or transferred by a user to another site.
For example, a user might request production of a specific DAOD format for their analysis based on a regular expression pattern.
If that expression is too general, DAODs might be produced for datasets that are never examined by the analysis team.
Similar issues can arise when DAOD formats are run together in a single job (called a ``train''): if a train is run on a dataset that is only needed in some of the DAOD output formats, the other DAOD formats will appear unused.
These issues have to be addressed by carefully manually going back through unused data, in each case understanding why it was produced, and tightening any rules or systems through which spurious data were created.

A common source of these unused datasets is connected to software issues in production.
If the physics output of a production is inaccurate, a new production might be required.
If an analysis team notices an issue after looking at a few datasets, without having looked at the rest, they might request a fresh production.
If the original production is not deleted, it will show up later as unused.
This can be the case when teams do not report data as invalid to the data management team, and instead simply request a new production and assume that the old data will be deleted in due course.
Again, such issues have to be identified, particularly when they result in large quantities of unused data, by systematic follow-up by production managers.

\subsection{Alternative resource usage}

ATLAS endeavors to make full use of all resources that it has been allocated.
For example, the high-level trigger farm, a cluster of CPU deployed at the experiment site for real-time data processing and event selection, comprises almost 2.3~MHS23.
This CPU farm is critical for operations, but when ATLAS is not taking data it can be used for processing of Monte Carlo simulation instead~\cite{SOFT-2022-02}, saving the experiment the equivalent of an additional 60~000 cores of compute.
Similarly, the CERN production site has an allocation dedicated to prompt processing of the data as it streams from the detector.
The site, known as the CERN Tier~0, is sized according to the need during peak collider operation.
Again, during a significant fraction of the year it is not needed for prompt detector data processing and can therefore be used for other workloads, saving the experiment the equivalent of about 75~000 cores of compute worldwide.

One aspect of the experimental policy has relatively poorly understood environmental implications: the use of HPC systems and commercial clouds in comparison to more traditional high-throughput computing (HTC) at WLCG sites.
The use of commercial clouds is still modest, around 1\% of the CPU time of the experiment, while the use of HPC systems is extensive, approaching one third when both pledged and unpledged CPU is included.\footnote{WLCG sites may ``pledge'' compute resources to ATLAS as an indication of what they expect to be able to provide in the coming year. Any extra resources if a site exceeds that pledge, and all resources from any computing sites that do not pledge resources are considered ``unpledged''~\cite{SOFT-2022-02}.}
The environmental impact of WLCG sites can be controlled to some degree, but the experiment has no control over the standard practices of commercial cloud providers.
Some of the standards, like the long-term use of computing hardware in WLCG sites, are not normally disclosed by commercial cloud vendors.
HPC systems normally have a relatively short lifetime of order five years, and the equipment is often then sent back to the vendor for reuse, recycling, or disposal.
On the other hand, commercial sites tend to procure renewable energy or have on-site sustainable power generation.
A detailed comparison of the environmental impacts of these systems would likely prove difficult due to non-disclosure agreements and proprietary information, but as more information becomes publicly available it may be worth undertaking.
Environmental considerations are unlikely to dictate more or less extensive use of these resources, as financial considerations are critical, but understanding the implications would be useful nevertheless.

\subsection{Open data}

Finally, it is important to consider the implications of some aspects of the experiment's computing policy on the sustainability of the field as a whole.
ATLAS has recently begun to release Open Data for research, along with a significant volume of Monte Carlo simulation corresponding to those data.
In addition, the experiment is planning to release billions of events, corresponding to hundreds of millions of CPU core-hours, of event generator output to the research community and general public.
Oftentimes, substantial effort is wasted and computing resources are consumed within the phenomenology community duplicating the event generation that ATLAS has already run.
Small groups tend to work in isolation, sometimes producing similar samples or simply trying to reproduce the samples that the experiment has already created.
By releasing these data, ATLAS can help reduce the environmental impact of the high-energy physics community at large~\cite{OpenEvgen}.


%
\section{ATLAS-specific site and user actions}
\label{sec:atlasactions}

While some hardware actions are generally beneficial and broadly applicable (see Section~\ref{sec:sites}), the particular policies and patterns applied within ATLAS computing offer opportunities for site administrators and developers to work together to reduce the environmental impact of the experiment.
These actions may not be generally applicable, but they offer many opportunities and pathways for other groups to follow to examine their usefulness.

\subsection{HEPScore and power measurement}
\label{sec:HS23}

To evaluate the computing power of resources, ATLAS employs the HEPScore benchmarking suite~\cite{HEPScore}.
This suite includes seven standard workloads from the LHC experiments, including ATLAS.
These workloads are chosen to provide a ``typical'' mixture of low-level instructions, memory loads, and other features so that the score is a more accurate reflection of a CPU's capabilities for ATLAS workloads than a standard benchmark like LINPACK~\cite{LINPACK}.
The HEPScore from 2023 (HS23 score) is important to understand which CPUs and configurations provide the best throughput per Watt, as well as the best throughput for the same cost.
It is also one of the key tools the collaboration can use to evaluate the performance of new platforms.
The benchmarking suite is intended to test CPU and memory combinations, and therefore to limit disk (input or output) and network usage.
For some workloads, particularly user analysis, this makes it less representative, but for the most time-consuming workloads the suite provides a fair characterization of resource use.

\begin{figure}[t!]
\centering
\includegraphics[width=\textwidth]{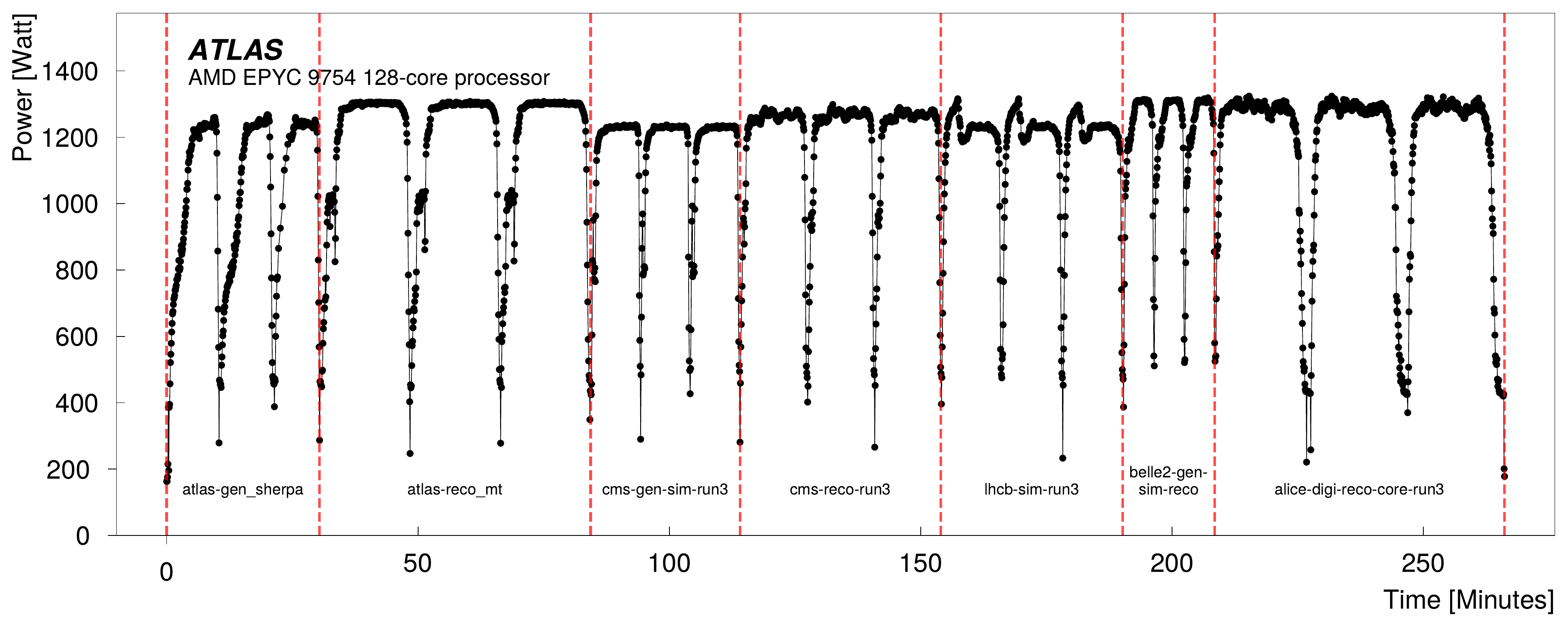}
\caption{The power consumption of a typical processor during the running of the HEPScore benchmarking suite. Each point represents a power measurement. Seven different sets of three runs of individual workloads chosen to be representative and described in Ref.~\cite{HEPScore} are distinguished with dashed vertical lines.}
\label{fig:HS23timeline}
\end{figure}

Considerable effort had to be put into understanding the power-use profile of the benchmark suite in comparison to standard production, to correctly calculate the score per Watt.
For example, the power consumption of a CPU, shown in Figure~\ref{fig:HS23timeline}, shows significant spikes and troughs during some workload initialization periods.
If the simple peak power consumption is used for the score, it will not correctly reflect typical production.
Similarly, the average is artificially low due to the frequent start/stop periods of the benchmark, which do not occur in normal production jobs that run continuously for 12 hours.
Various metrics were examined, and the most stable and simple to calculate metric is ``q85'', the 85th-percentile power measurement.

\begin{figure}[t!]
\centering
\subfloat[]{
\includegraphics[width=\textwidth]{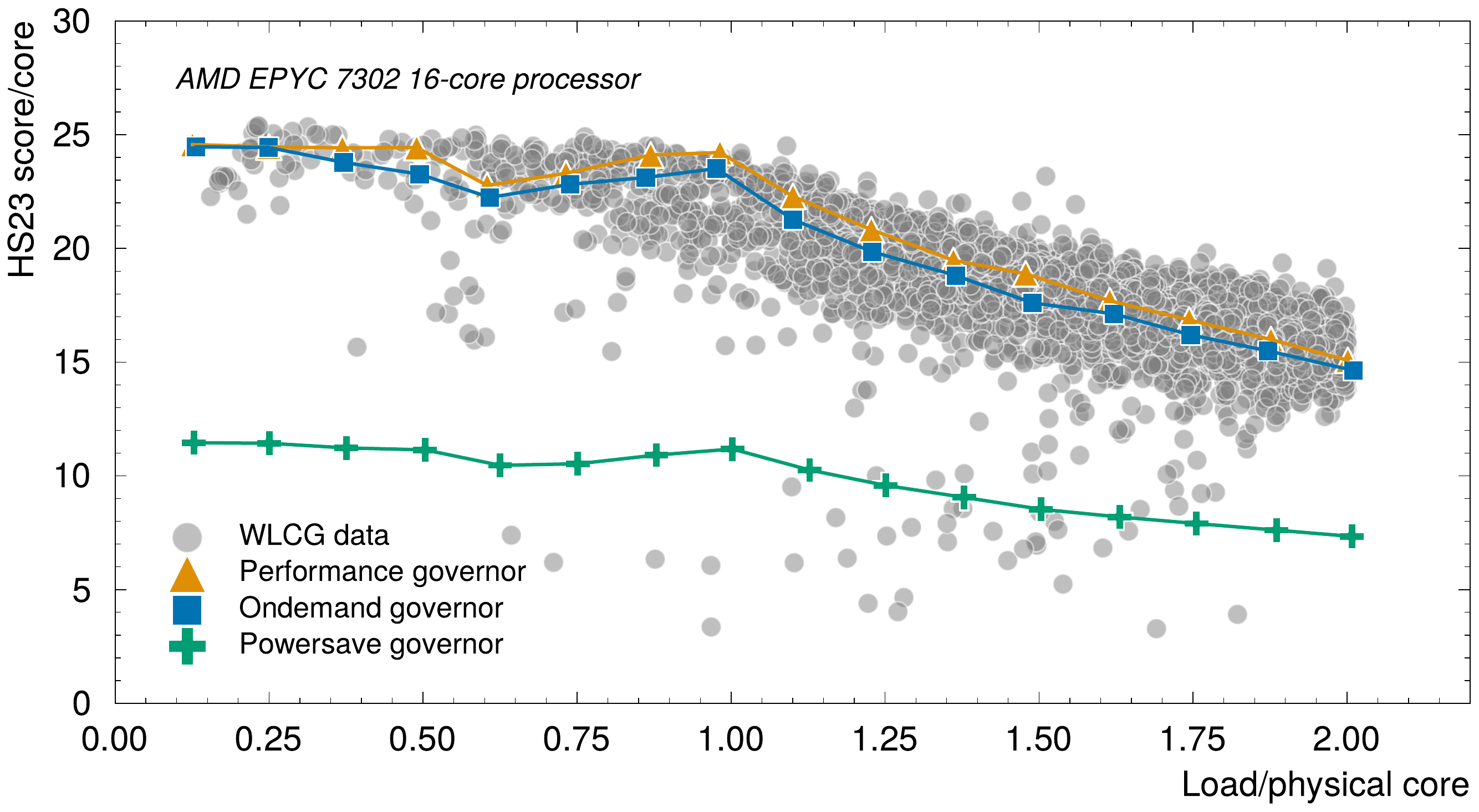}
} \\
\subfloat[]{
\includegraphics[width=\textwidth]{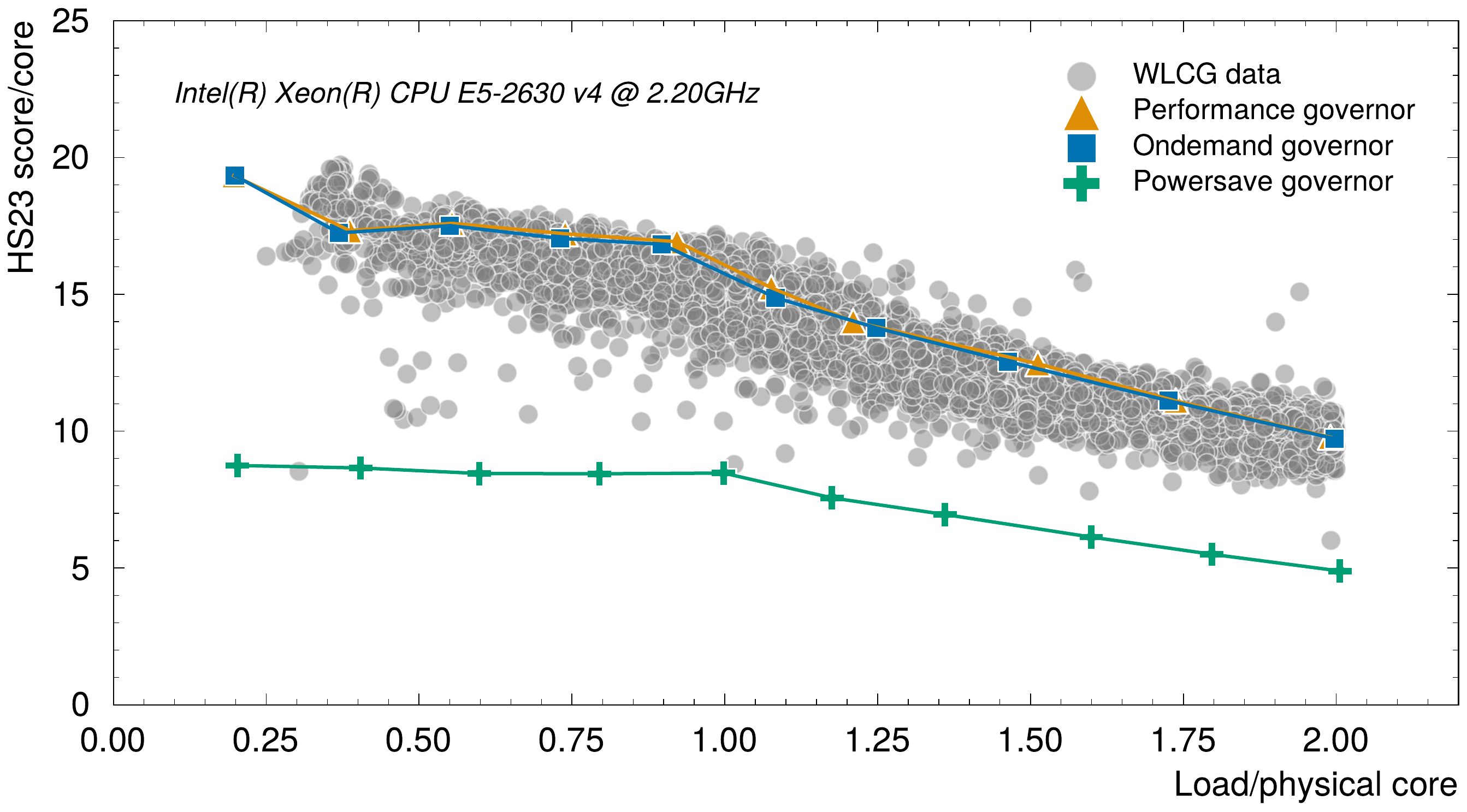}
}
\caption{The evolution of HS23 score with CPU load for (a) AMD and (b) Intel CPUs. The gray points show data from the WLCG sites, while the colored series connected with lines indicate specific choices of frequency governor on benchmark machines. The data are from Ref.~\cite{HEPScoreSustainability}.}
\label{fig:governors}
\end{figure}

The load on the processor --- how much work is running simultaneously --- can also affect its performance in ways that depend on the workload.
The effectiveness of hyperthreading, for example, in improving the performance per Watt depends on the workload and its use of the CPU.
There are also a variety of power regulation mechanisms in modern CPUs that can affect the scaling of power consumption with load.
Data from the WLCG sites, shown in Figure~\ref{fig:governors}, reveal two distinct scaling regimes in the evolution of HS23 score with load.
These patterns are tied to specific configuration factors, such as turbo boost settings and thermal shielding, which can vary depending on the CPU brand and generation.

To investigate these aspects more thoroughly, dedicated, isolated tests were conducted on bare-metal servers (i.e.~not running inside of containers) using different CPU frequency governors.
These governors set the policy for automatically changing the CPU frequency during operation to improve performance or to save power, depending on the goals of the user.
The benchmarking was performed on AMD EPYC 7302 16-Core CPUs and Intel\textregistered{} Xeon\textregistered{} E5-2630 v4 \@ 2.20~GHz CPUs.
The number of threads was scanned over multiples of four, to approximate the behavior observed in the WLCG site data.
The results are shown in Figure~\ref{fig:governors}, using data from Ref.~\cite{HEPScoreSustainability}.
Three governors were tested: ``performance,'' ``ondemand,'' and ``powersave.''
For both CPU models, the powersave governor consistently produced about 50\% lower performance results, while only reducing power consumption by about 45\%.
The HS23 per Watt, therefore, is lower when using the powersave governor.
Examining the data from the WLCG sites, most servers align with either the performance or ondemand characteristics --- two modes that tend to overlap.
Notably, the performance governor prioritizes maximum output regardless of power cost, which explains its consistently slightly higher scores than the ondemand governor.
The precise behavior varies depending on many factors including turbo boost settings, CPU thermal shielding, and CPU model.
This observation underscores how performance can be directly influenced by CPU configuration and governor choice.
In environments where energy efficiency is paramount, the powersave or ondemand governors may offer the best trade-off, depending on the specific needs and workloads.

These studies are also key to the evaluation of new hardware platforms.
The recent adoption of ARM processors within ATLAS computing was the result of significant work motivated by both financial and power usage considerations~\cite{ARMproc}.
ATLAS computing has also demonstrated the use of commercial cloud resources for development of new platforms~\cite{Megino:2024htf,SOFT-2023-02}.
Using cloud resources allows the collaboration to carefully evaluate new platforms, run productions at modest scale, and perform detailed validation without having to purchase resources that might be wasted if the validations are unsuccessful or the resources prove more costly than existing options.
Eventually, to get detailed power monitoring information, it is currently necessary to procure resources.
If in the future commercial cloud providers offer better access to this information, it may be possible to evaluate new platforms entirely in the cloud before purchases have to be made.

Understanding the power usage of CPUs across generations is also critical to understanding the optimal approach to hardware retirement.
The issue is still not well understood: whether it is better to run old hardware and amortize the embodied carbon over a longer period, or to retire and donate or recycle old hardware to use only new, more power-efficient CPUs.
The optimum depends on the relative importance of operational and embodied carbon, and therefore on the carbon intensity of the power grid where the site operates.
WLCG sites tend to run hardware well past its warranty lifetime; a modest fraction ($<10$\%) of CPUs running ATLAS software are more than 10 years old (see for example Refs.~\cite{SOFT-PROC-2023-001,WLCG-Energy}).
Often site administrators extend the lifetime of hardware by scavenging parts from failing nodes, and in some cases hardware is donated to groups running less resource-hungry workloads.
Several efforts are underway to understand the issue of optimal hardware lifetimes in more detail (see e.g.~Ref.~\cite{LifeCycle}).

\subsection{Power variation and frequency adjustment}

\begin{figure}[pt!]
\centering
\includegraphics[width=\textwidth]{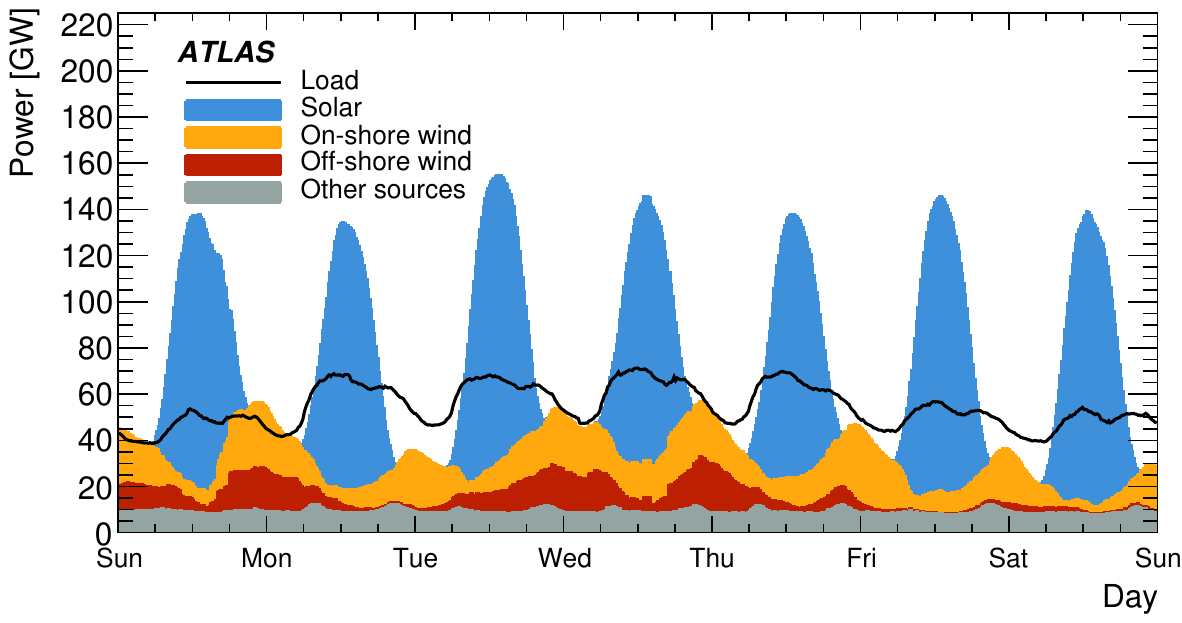}
\caption{An extrapolation of renewable power sources in Germany and power demand to 2030, based on Refs.~\cite{German-Power,Fraunhofer}. The load (demand), on-shore and off-shore wind, and solar power production are scaled based on data from an example week at the end of May, 2023 and projections for consumption and renewable energy construction projects. Other sources include hydroelectric, biomass, geothermal, and waste-to-energy production.}
\label{fig:Rod}
\end{figure}

One issue that can be addressed through collaboration between ATLAS site administrators and software developers is the variation in power grid availability and carbon intensity.
The variation of carbon intensity is visible already in many regions: in certain periods of the day, renewable energy is more broadly available, and the carbon intensity of the power grid is lower.
Other periods see reduced production of renewable power, particularly solar and wind.
The load on the power grid also varies during the day.
As the renewable power sources expand over the coming years, it is expected that during peak production periods the power production from renewables will significantly outstrip the demand.
This is shown in Figure~\ref{fig:Rod}, for the case of modeled projections for Germany in 2030, based on data from Refs.~\cite{German-Power,Fraunhofer}.
There are several ways in which sites can take advantage of this variation, by increasing production when renewable energy is abundantly available and might even be curtailed if not used, and reducing production and therefore power load when it is not.

\begin{figure}[pt!]
\centering
\includegraphics[width=0.8\textwidth]{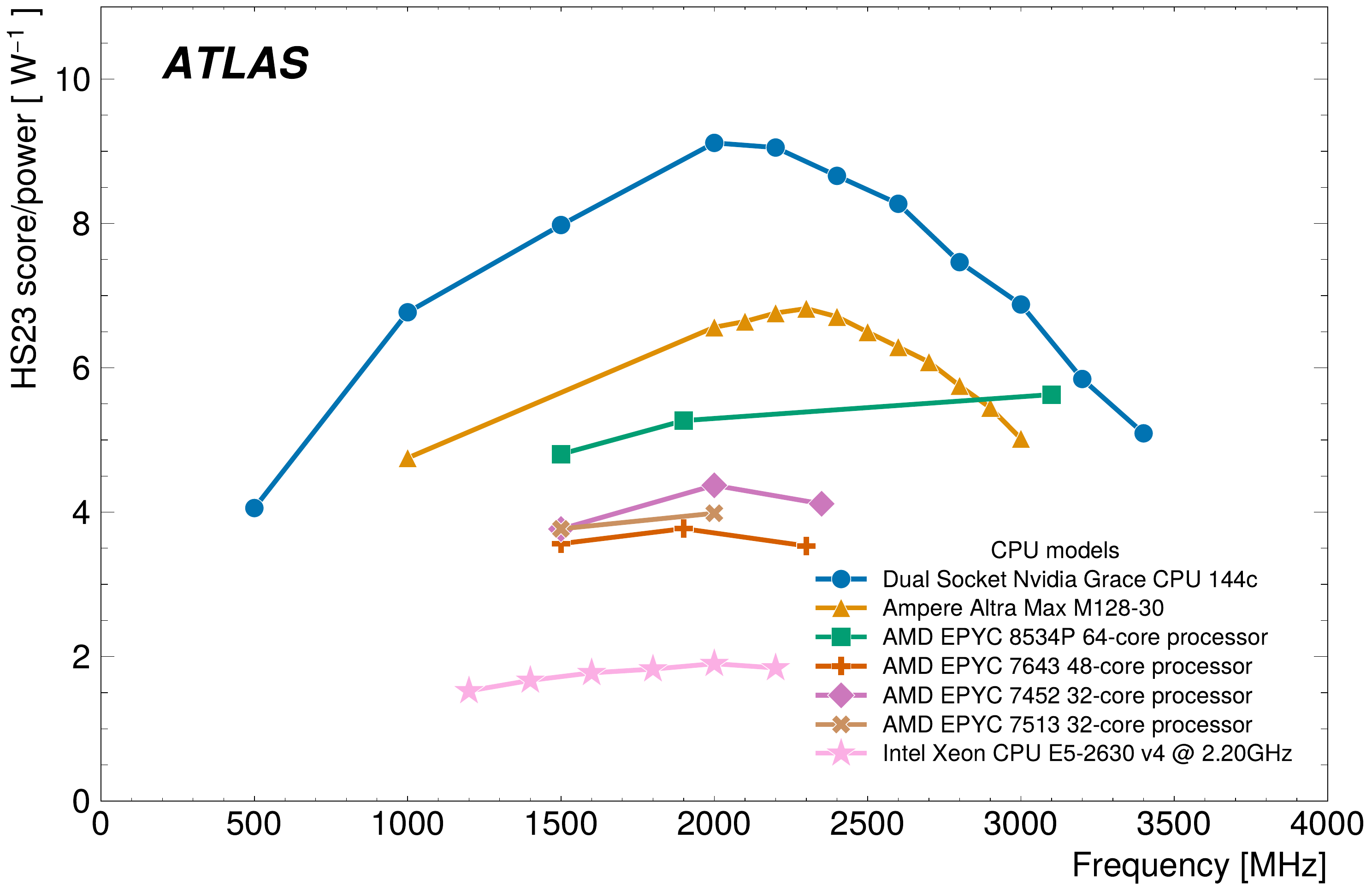}
\caption{The HS23 score (a proxy for throughput) per Watt as a function of the processor frequency for seven CPU models. The 85th percentile power consumption during the benchmark running is shown.}
\label{fig:CPUfrequency}
\end{figure}

Modern CPUs have controls for frequency adjustment.
Lower frequencies generally result in reduced work and reduced energy use, but because many ATLAS jobs are restricted by memory access patterns rather than simply CPU cycles, the reduction in throughput is less than linear.
Additionally, the power usage of a CPU is proportional to frequency and voltage squared, and when the CPU frequency is reduced the voltage is also reduced.
Both of these effects result in the throughput per Watt being significantly better for ATLAS jobs on CPUs with reduced frequency.
This improvement can be seen in Figure~\ref{fig:CPUfrequency}, showing the HS23 score (as a proxy for throughput) per Watt for various processors as a function of CPU frequency.
In all cases, running the CPU below its normal design frequency is beneficial to improving the HS23 score per Watt.
The improvement and best set point depends on the CPU model, and therefore must be tested with each procurement.

CPU scaling offers a straightforward way for site administrators to immediately increase the throughput per Watt of their site.
Although there is an improvement in HS23 per Watt, the total throughput per second of the site will be reduced.
To provide the same throughput per second, more CPUs will need to be purchased.
The optimal working point balances the power costs against the cost of additional hardware, and balances the reduction of environmental impact from reduced power consumption with the increased embodied carbon from the additional hardware.
This optimum is site and time dependent and must be more carefully examined for future purchases.
The same arguments apply to GPU frequency settings as well~\cite{GPUlimits}.

In extreme cases, the variation in power grid availability can result in brownouts (drops in power grid voltage or temporary blackouts) or blackouts of sites, whether scheduled or announced on relatively short notice.
In these cases, the site may be taken completely offline.
Currently, sites must either ``drain'' all processing completely or kill all running jobs before such a shut down to avoid leaving many tasks in a corrupted state.
ATLAS grid jobs typically run 12--24 hours, with tails on some sites up to 3--4 days.
To completely drain a site, no new work can be submitted for at least that period, which can cost hundreds of thousands of CPU-hours for a large site.
To avoid the loss of processing, checkpointing (i.e.~taking a snapshot of the current state of a running program, storing it on disk, and later restoring the running program from that snapshot) could be implemented within the software.
Several ATLAS production workloads are already capable of checkpointing, but the functionality has not been built into the production system to allow sites to send appropriate signals to the jobs, reload their states, and correctly report the job status and resource usage in the monitoring systems.
This additional work is non-trivial but could be advanced in preparation for the HL-LHC to improve site operations and efficiency.
Checkpointing has other applications as well.
For example, if a site is to be taken offline for a regular upgrade, the CPU can continue to run until just before the scheduled upgrade and can be promptly brought back into a running state after an upgrade.
If the use of checkpointing produces inconsistencies in the system that causes all checkpointed jobs to fail, the loss of CPU will be no worse than the current system.

Several sites, including commercial cloud providers, are looking into alternative powering mechanisms to resolve the renewable power availability issues.
One outstanding question is whether the development of battery systems, currently used for emergency power (e.g.~UPS systems) will develop substantial additional capacity in the coming years, such that they could be used to power the entire data center during periods of high-carbon-intensity power production, and then could be recharged during periods of very low-carbon-intensity power availability.
Such a battery system could have significant embodied carbon, but might be able to substantially reduce the operational footprint of the site.
Similarly, some commercial providers are looking into miniature nuclear reactors to provide local low-carbon energy sources~\cite{Minireactors}.
In all cases, the commercial development of these options will significantly outpace HEP-specific developments, and ATLAS sites can take a wait-and-see approach to the technological advances.

Two additional advantages of ATLAS computing in comparison to many other sites and operations are consistency and latency.
CPUs are kept busy at all times when they are available, including CPUs provided opportunistically to the experiment beyond its standard resources.
Excluding resources based on sustainability considerations would impose financial costs that the collaboration could not afford today.
The consistency means that resources can be scoped for typical use, with a small operational safety margin, rather than being scoped for a peak well above typical use.
The acceptance of latency means that if demand is high, generally work can be prioritized rather than requiring that resources are expanded temporarily.
In the case of networking equipment, for example, the continuous stream of transfers around the world means that their power consumption is steady, and the embodied carbon is only what is required by the typical use pattern and safety margin.


%
\section{General site actions}
\label{sec:sites}

Many actions that sites can take towards sustainability and reducing environmental impact are not specific to ATLAS, or even to HEP.
Groups worldwide are investigating these issues (see for example Refs.~\cite{EEHPCWG,GreenDisc,GreenDIGIT,IdlePower}).
Several of these groups have close connections to the ATLAS experiment: some have used ATLAS sites as examples or case studies; in other cases ATLAS site teams have directly been involved with the studies.
Specific examples are included in this section to do with storage optimization, data center construction, and cooling.
Many general recommendations are parameterized in terms of common operational parameters like average load and hardware lifetime, and optimal working points may be site- or region-specific.

Several ATLAS sites are also investing effort in broader examinations of power and environmental impact studies.
Site power profiles vary substantially depending on the site, its capabilities, and its configuration.
Storage often makes up 25\%--45\% of the total site power consumption.
Where tape storage is provided, the system normally consumes another 10\% of the site's total power.
The remainder is primarily consumed by CPU, with a very small fraction for additional services like networking elements (see, for example, Ref.~\cite{CarbonQuanDRI} for a UK study).
In Germany there are several ongoing efforts to understand and reduce the carbon footprint and environmental impact of computing centers (see for example Refs.~\cite{Germany1,Germany2}).
These efforts are now being harmonized across the WLCG, where dissemination of best practices and standards has been given the highest priority (see for example Ref.~\cite{WLCGWorkshop}).

One example of a generic site action that was motivated by ATLAS studies is the introduction of more flexible memory allocation in site queues.
Most sites provide 2~GB of memory per CPU core; historically, jobs requiring 4~GB of memory were assigned two CPU cores, and one of the cores would sit idle.
By introducing a greater range of memory requirements in queues, ATLAS can submit low-memory work (e.g.~requiring only 1~GB of memory) and maintain the 2~GB per core average while still running a substantial number of higher-memory jobs.
This reduces the amount of idle CPU at the site that draws power but does not produce useful work.

The ATLAS WLCG sites are more directly under the supervision of the experiments.
They are generally responsive to recommendations and requests from ATLAS, and the administrators are collaborative and deeply connected to the experiment.
The diversity of ATLAS resources, however, means that not all sites providing resources are so well integrated.
About one third of ATLAS CPU is provided by HPC systems and volunteer resources, many of which were constructed to serve communities and use-cases other than high-energy physics.
While the experiment can still offer recommendations and suggest actions to these sites, a broader community must come together to influence them.


\subsection{Storage optimization}
\label{sec:storage}

%

%

The carbon footprint of storage systems remains an understudied area, with relatively few papers focused on reducing Scope 2 (operational) ~\cite{Storage_rao2024,GreenMatch2018,Cadre_xu2015,GreenCassandra_katsak2015} and Scope 3 (embodied)~\cite{GreenSKU_wang2024,CarbonHPC_li2023,EmbodiedSSD_tannu2023} emissions of storage systems.
Recent studies highlight the significant embodied carbon footprint of storage systems, contributing approximately 21\%--42\% to the data center's embodied carbon footprint~\cite{CarbonHPC_li2023}. Strategies like extending storage device lifetimes or using denser devices require further investigation to assess their trade-offs with operational carbon emissions. Research on reduction of the operational carbon footprint has primarily concentrated on shifting of foreground and background tasks to use low-carbon power opportunities.

\begin{figure}[ht!]
\centering
\subfloat[]{
\includegraphics[width=0.47\textwidth]{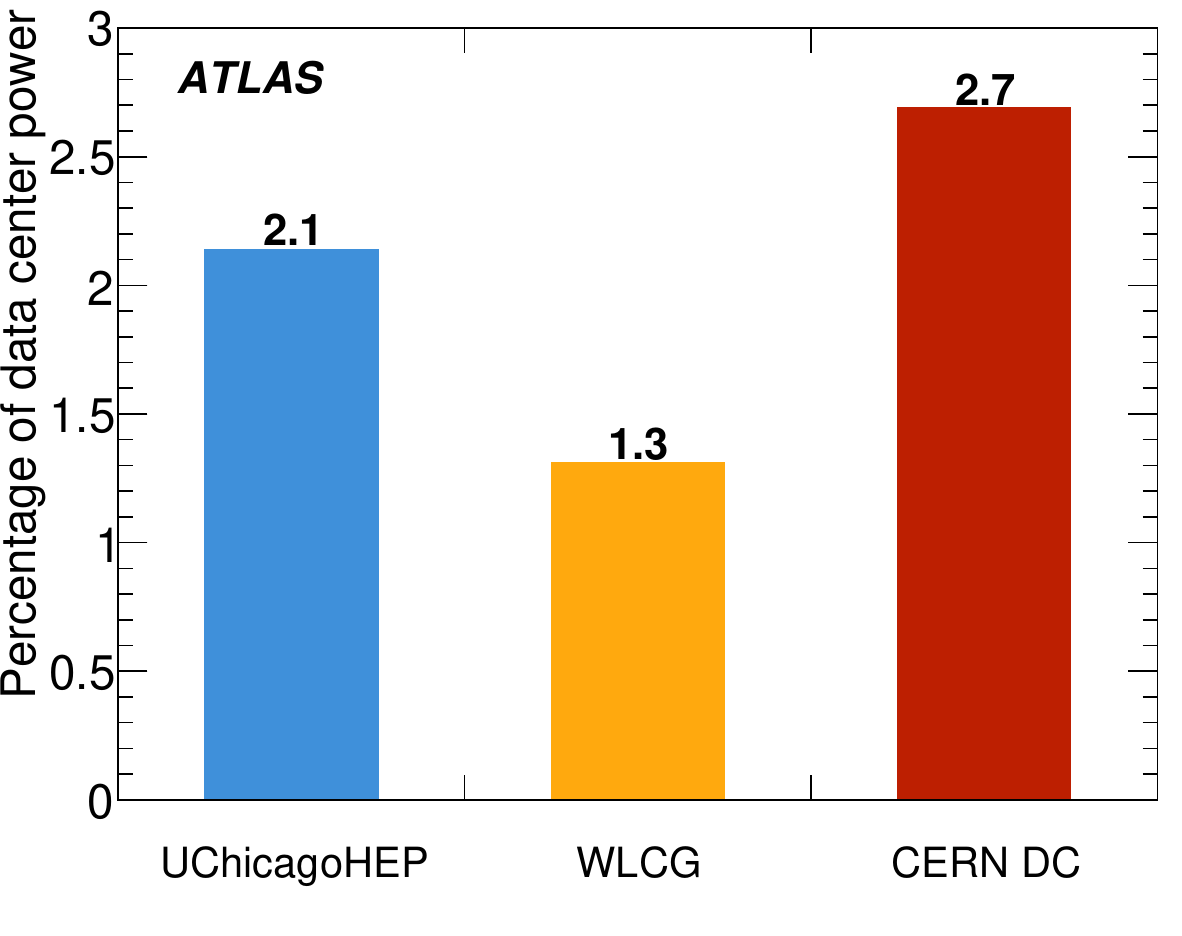}
}
\subfloat[]{
\includegraphics[width=0.47\textwidth]{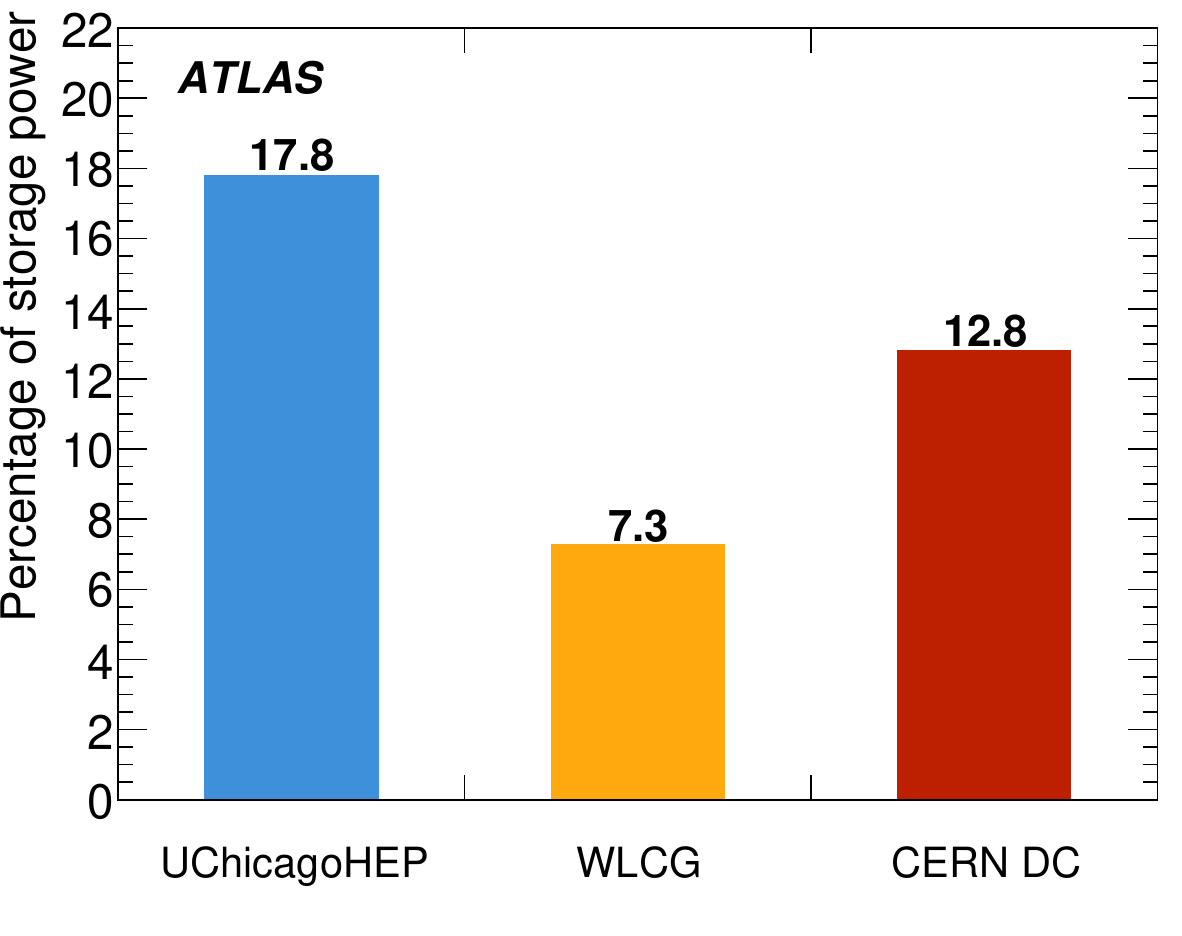}
}
\caption{Power for background tasks as a percentage of (a) data center power and (b) data center storage power, for three HEP site scenarios.}
\label{fig:bg-power-relative-dc-storage}
\end{figure}

The operational carbon footprint of storage systems can be reduced by temporal shifting of background tasks to low-carbon periods. Background tasks such as deduplication, backup, and scrubbing are essential for maintaining storage reliability and efficiency. These tasks usually handle large volumes of data and are deferrable. In Ref.~\cite{Storage_rao2024}, a workload model for background tasks was built to understand the I/O, compute and power costs of these tasks. In this paper, the background tasks workload model is applied to three HEP site scenarios that follow available site configurations: the Tier-0 (the CERN data center, ``CERN DC''), a large Tier-2 site (University of Chicago HEP, ``UChicagoHEP''), and all sites worldwide (``WLCG''). These tasks consume 1.3\%--2.7\% of the overall data center power and 7.3\%--17.8\% of the storage power, as shown in Figure~\ref{fig:bg-power-relative-dc-storage}. The difference between results can be attributed to variations in the amount of storage relative to compute, storage configuration and management policies, which substantially influence the power consumption of background tasks. For instance, the storage capacity of WLCG and the CERN data center is 45--50 times greater than that of the Tier-2 site. Tasks like data scrubbing, which process all stored data, constitute a significant portion of background task power consumption. Many background tasks also do not scale linearly with storage volume, which contributes to the observed differences between sites~\cite{Storage_rao2024}.

The operational carbon footprint of the background tasks in the CERN data center scenario is about $49.4 \times 10^3$~kgCO$_2$eq/year. The CERN data center is powered by the French power grid, where nuclear power is the dominant source of electricity generation and therefore a low average carbon intensity (ACI) is typical. Here, the fluctuations in ACI are lower compared to power grids where renewable energy is the dominant power source. Several other power grids within the United States are considered for modeling the variation in carbon intensity from different power sources: CAISO (California), MISO (Midwest and South), ERCOT (Texas), and SPP (Central South). The carbon emissions of background tasks for a site like the CERN data center would be 3.5 times larger in CAISO and 8.2 times larger in MISO than in France. This is due to different power generation sources. For example, in CAISO, solar energy is the predominant renewable energy source. The ACI is lowest during daylight hours when solar generation peaks. These fluctuations in ACI present an opportunity to reduce the storage system's operational carbon footprint by scheduling background tasks during periods of lower carbon intensity.

\begin{figure}[htb]
\centering
\subfloat[]{
\includegraphics[width=0.9\textwidth]{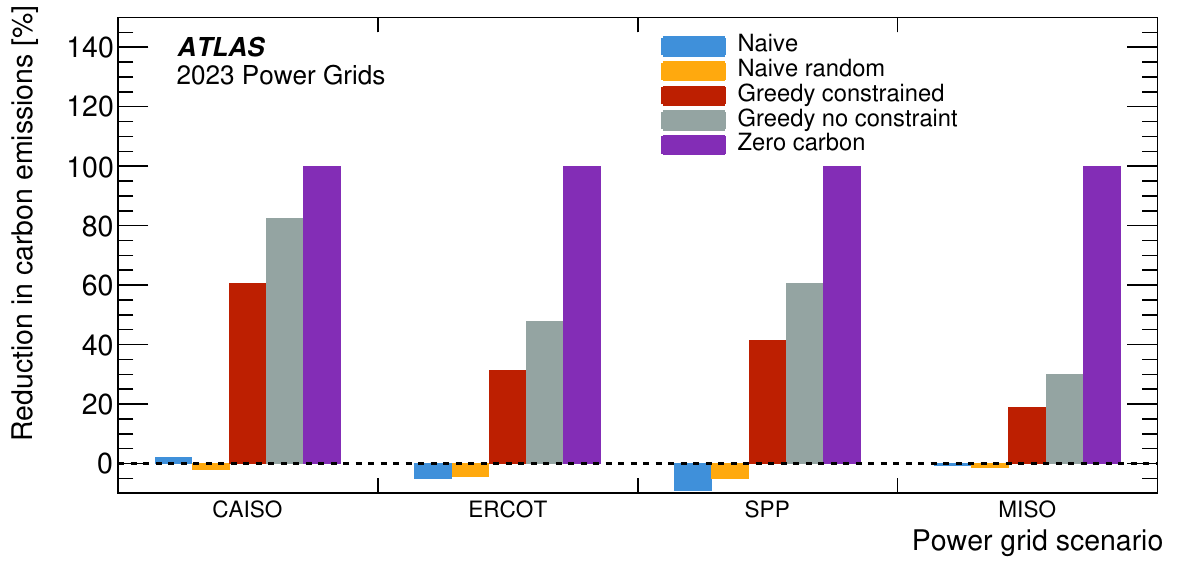}
} \\
\subfloat[]{
\includegraphics[width=0.9\textwidth]{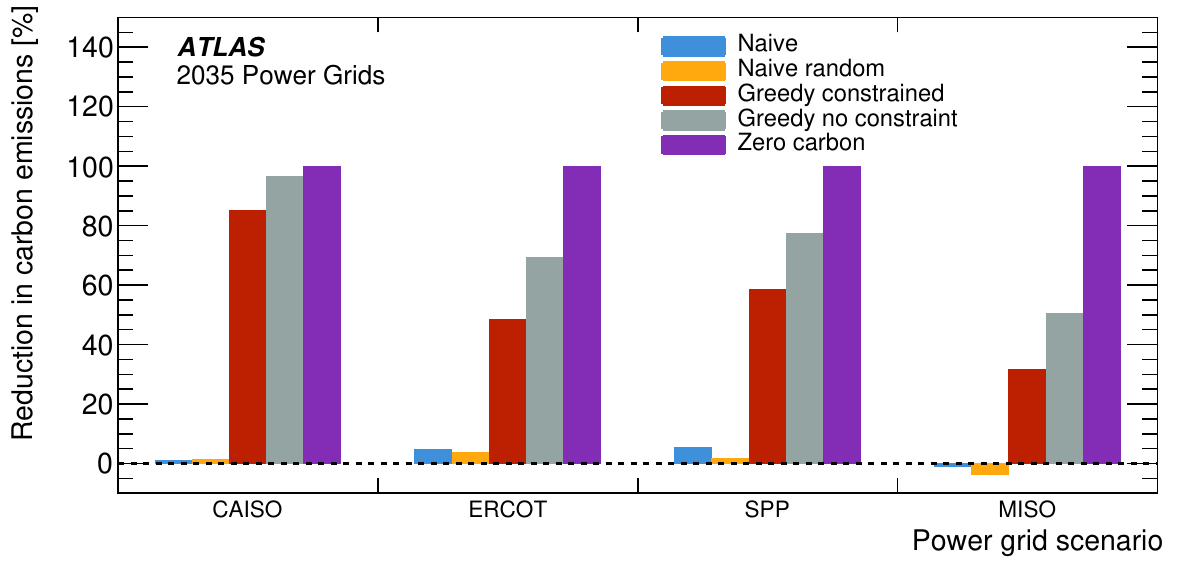}
}
\caption{The percentage reduction in carbon emissions of background tasks using several approaches to scheduling, for the CERN data center, for (a) the power grid models of 2023 and (b) the power grid models of 2035. The various power grid models and scheduling strategies are defined in the text.}
\label{fig:policy-eval}
\end{figure}

Figure~\ref{fig:policy-eval} shows the percentage reduction in carbon emissions, calculated using ACI values, relative to the baseline emissions for the CERN data center site scenario based on current scheduling policies, for the aforementioned power grid scenarios. The evaluation is based on the methodology described in Ref.~\cite{Storage_rao2024}. ``Naive'' denotes the usual configuration of scheduling background tasks around midnight. ``Naive Random'' represents scheduling tasks at random times of day. The results of these policies depend on the ACI value at the task running time and increase carbon emissions in some scenarios. ``Greedy Constrained'' optimizes for carbon while scheduling tasks within a constrained time period (e.g.~the tasks must run daily). It achieves a carbon emission reduction of  60.6\% in CAISO and 41.3\% in SPP. With no task period constraints, ``Greedy No Constraint'' achieves up to 82.4\% reduction of emissions in the CAISO power grid scenario. This is closer to the ideal scenario of ``Zero Carbon'', where tasks are scheduled only during zero carbon periods. In 2035, with the rapid decarbonization of power grids~\cite{AIInferenceFuture}, the variation in ACI will increase, and these policies achieve greater reductions in emissions relative to the naive policies, up to 96.4\% in the CAISO power grid scenario.

In short, the temporal shifting of background tasks to low carbon periods is an effective approach to reducing the operational carbon footprint, as estimated using the ACI, of storage systems. It can achieve a decrease in the carbon emissions of data center storage of up to 7.8\% with the ``Greedy Constrained'' approach and 10.6\% with the ``Greedy No Constraint'' approach.

Storage configurations are highly site-specific, and one of the other ongoing efforts is to understand whether there are opportunities for configuration improvements that might reduce the environmental impact of the sites with minimal operational disruption.
Simple differences in the redundancy of storage configurations (RAID levels) and drive layouts can produce significant changes in the power per TB of available storage.
Similarly, many sites have reported that the CPU load on disk servers is relatively low.
Some therefore run additional production jobs on CPU server head nodes to take advantage of the additional computing power~\cite{CERNBeer}.
Although it is unclear whether these studies and actions will generalize to non-HEP workloads, they have already proven useful when applied to LHC or ATLAS sites.


\subsection{Data center construction}
\label{sec:building}

The carbon footprint of construction is a complex issue, with many variations in estimates of supply chain carbon in particular (see, e.g., Ref.~\cite{ConstructionComplexity}). Particularly in countries with relatively clean energy grids, embodied carbon (construction carbon) may represent a significant fraction of the total footprint of a building, even for office buildings with 50-year life expectancy (much longer than a typical data center)~\cite{ConstructionFinland}. The primary source of carbon during construction is the sourcing of materials~\cite{ConstructionReduction}, with significant opportunity for the improvement of the overall carbon footprint. The machinery used to gather and supply the primary materials, as well as those used to construct the building, also contribute significantly to the total footprint, as these are often difficult to power with green energy~\cite{ConstructionChina}. The materials most commonly used in data center construction, steel and concrete, are among the most expensive in terms of carbon footprint.

Many studies were done of the total carbon footprint of building construction, for a wide variety of construction projects~\cite{ConstructionSurvey}. Steel-reinforced concrete buildings, like many data centers, typically have carbon footprints of 200--800 kgCO$_2$eq/m$^2$~\cite{ConstructionSurvey,CircularEcology}. Several efforts to characterize and reduce the total carbon footprint of data center construction have been undertaken, including LEED data center certification from the U.S. Green Building Council~\cite{LEED}.

When constructing a new data center, the main consideration is the change in PUE. This change leads to a simple relationship for the amortization of a building by comparing the total carbon footprint of the construction of the building to the footprint of the power consumption of the building. The carbon saved per year can be written as:

\begin{equation}
\frac{\Delta \textrm{PUE}}{\textrm{PUE}_\textrm{old}} \times \textrm{Power} \times \textrm{GCI},
\end{equation}

where $\Delta\textrm{PUE}$ is the change in the PUE from the old to the new building, $\textrm{PUE}_\textrm{old}$ is the PUE of the old data center, $\textrm{Power}$ is the annual power draw of the old building, and $\textrm{GCI}$ is the power grid carbon intensity. This can be compared to the total building construction carbon footprint to estimate the number of years before a new data center ``pays off.'' This assumes that the computing hardware is simply moved from the old building to the new building rather than buying entirely new hardware, and that the carbon cost of transportation between the two buildings is small (generally true because of the short distances between the two). In some cases, particularly for significant expansions of computing resources, entirely new computing hardware is purchased. This misses the opportunity for carbon savings, but often is justified by other considerations (e.g.~the importance of hardware uniformity in the new data center).

Typical large modern data centers include around 10~000~m$^2$ of floor space. The GCI varies significantly by country; for French power in 2024 (similar to what CERN uses) generation cost 0.033~kgCO$_2$eq/kWh; for California in 2024 it was about 0.227~kgCO$_2$eq/kWh, typical in the USA~\cite{ElectricityMaps}. Power draw for a large modern data center is typically around 5~MW. Some examples of annual carbon savings for a new data center are:

\begin{itemize}
\item For a 5~MW data center in France, changing the PUE from 1.6 to 1.1:
\begin{equation}
\frac{0.5}{1.6} \times 5000~\textrm{kW} \times 8760~\textrm{h/year} \times 0.033~\textrm{kgCO$_2$eq/kWh} = 452~000~\textrm{kgCO$_2$eq/year}
\end{equation}
\item For a 5~MW data center in California, changing the PUE from 1.2 to 1.1:
\begin{equation}
\frac{0.1}{1.2} \times 5000~\textrm{kW} \times 8760~\textrm{h/year} \times 0.227~\textrm{kgCO$_2$eq/kWh} = 829~000~\textrm{kgCO$_2$eq/year}
\end{equation}
\end{itemize}

For a 10~000~m$^2$ data center, a 500~kgCO$_2$eq/m$^2$ carbon cost, the middle of the range above, gives a total construction carbon footprint of $5\times 10^6$~kgCO$_2$eq. For the French example, this suggests a total amortization time of about 11 years. For the California example, this suggests a total amortization time about about six years.

As the grid steadily de-carbonizes, the carbon cost from operation will be reduced; although goals are lofty, one can expect that this is unlikely to reach a factor of two in time for the HL-LHC, particularly considering global evolution. However, there are significant opportunities to reduce the carbon cost of new construction~\cite{ConstructionReduction}. In essence: if a new data center can be brought online with a significant PUE reduction, even of order $0.1$ in regions where the GCI is high, it is very likely to be worthwhile in terms of the total carbon footprint of computing. Similarly, any renovations that make significant change to the PUE are likely to have a rapid net benefit on the total carbon footprint of the data center.


\subsection{Cooling systems}
\label{sec:cooling}

The PUE of a data center is tightly coupled to the design of the building, as well as the type of cooling systems used both in the building and for the computing. The most common options for computing are currently air cooling, using fans and pressure differentials to force air flow around the computing units, and water cooling, circulating cool water throughout the cooling racks to cool the computing. Various more exotic cooling systems, including liquid immersion cooling, are being investigated worldwide, but these tend to have more significant capital costs and maintenance issues that make them less desirable for standard WLCG sites.

Liquid (water) cooling systems are generally much more efficient than air cooling systems, and they can cool racks with significantly higher total thermal design power. There are significant capital costs associated with the transition to liquid cooling, but these can be overcome by cost savings from reduced power needs and careful planning~\cite{EEHPCLiquidCooling}. Many individual examples are available, most indicating some significant savings in power from the transition to liquid cooling~\cite{WasteHeat,Bytes2Heat}. For modern buildings that are designed with energy efficiency in mind, where the climate is temperate, and therefore the data center PUE is quite low, the transition from air to water cooling may not result in a significant savings. However, for older data centers with higher PUE, or in places with warmer climates even for part of the year, the transition to water cooling may result in quite significant carbon and cost savings. Staged transitions to liquid cooling solutions can also be beneficial. For example, direct liquid cooling with perimeter computer room air handlers can save 15\%--20\% of the total data center energy consumption compared to air cooling alone. Moving from perimeter computer room air handlers to close-coupled air-cooling can save an additional 15\%--20\% of the data center energy consumption~\cite{CoolingComparison}.

The transition from air to water cooling has some ancillary benefits as well. For example, water-cooled systems tend to be much quieter, and the lack of a ``hot aisle'' makes them much more pleasant to access and maintain. Air-cooled systems often run around 30\degree~C, making the reuse of heat from these systems considerably more difficult and expensive (see also Section~\ref{sec:wasteheat}). The transition to water cooling seems to have no significant down-sides, and should be undertaken wherever practical.

A transition from air to liquid cooling is being explored at one of the largest ATLAS Tier-1 sites, hosted at Brookhaven National Laboratory. The pilot project also includes the re-use of data center heat for buildings and domestic hot water. The eventual deployment of liquid cooling, after optimization and heat re-use, is expected to save 700~MWh of energy and 150~000~kgCO$_2$eq/year~\cite{ImranCooling}. In these cooling systems, there is significant scope for optimization by adjusting temperature set-points.

\begin{figure}[ht!]
\centering
\includegraphics[scale=0.6]{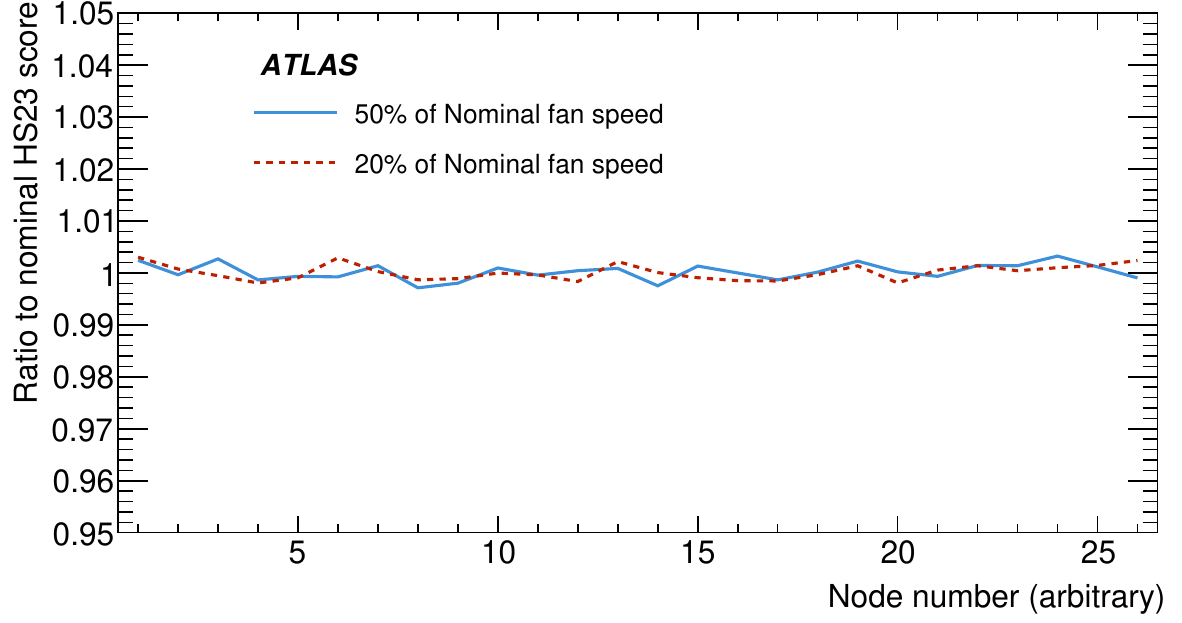}
\caption{Ratio of the modified HS23 score to the nominal for a selection of data center nodes, for several fan heat exchanger operating speeds relative to the nominal.}
\label{fig:fanspeed}
\end{figure}

Generally speaking, running a warmer system requires less energy for cooling. However, modern CPUs include complex temperature regulation functionality that automatically lowers the frequency to avoid overheating. One concern, therefore, is that a higher temperature set-point would result in lower data throughput. To investigate this issue, the HEPScore benchmarking suite was run on a test rack with 26 Dell x86 compute nodes, where rear-door heat-exchanger fan speeds were varied from 20\% to 100\% of normal. With the lowest fan speed the ambient air remained just below the maximum threshold defined by the CPU manufacturer of 35\degree~C. As shown in Figure~\ref{fig:fanspeed}, in all configurations the HS23 score was identical, implying the higher temperatures had no significant impact on data throughput. By raising the data center cooling set point by 10\%, which allows temperatures of all compute nodes to increase, a 15\% savings in energy can be achieved without compromising computing throughput.

\subsection{Waste heat usage}
\label{sec:wasteheat}

One of the few opportunities for computing to truly save carbon is in the reuse of waste heat. The computing systems generate significant heat when running. Historically, this heat was expelled into the atmosphere or, in colder climates, used to warm the data center itself, particularly offices associated with the data center. More recently, systems have been designed to capture waste heat and put it to use for warming buildings or for warming water to be distributed more broadly, particularly using liquid cooling~\cite{OpenComputeHeatReuse}. This is a practical solution only for data centers that are in reasonably well-populated areas; the transmission of waste heat over long distances may require significant time to be of value, particularly after accounting for the embodied carbon of the additional infrastructure. Still, data centers can provide a stable, year-round heat source, and the heat exchange systems can be quite efficient, recapturing 80\% of the heat generated by the computing systems~\cite{Aquasar}. In some regions of the world, heat reuse, or at least the exploration thereof, is required by law for new data centers; grants are also often available to assist with the infrastructure costs~\cite{OpenComputeHeatReuse}.

The carbon savings from waste heat usage depends on several factors, for example:

\begin{itemize}
\item The local environment, and how much heat is needed domestically or for local office buildings;
\item The typical local sources of heat, and how carbon-intense those sources are (e.g.\ an area heated by coal-fired plants would see much larger savings than an area heated by geothermal activity);
\item The efficiency of the heat exchange system, which depends on the cooling system used for the data center.
\end{itemize}

One way to quantify the impact of waste heat usage is through a modification of the PUE, sometimes also referred to as the Energy Reuse Efficiency (ERE), defined as~\cite{WasteHeat}:

\begin{equation}
\textrm{ERE} = \frac{\textrm{Building Power} - \textrm{Extracted Energy}  }{\textrm{IT Power Consumption}} = \textrm{PUE} - \frac{\textrm{Extracted Energy}}{\textrm{IT Power Consumption}},
\end{equation}

\noindent where the Extracted Energy represents a correction to the PUE and assumes that the carbon cost for heat generation is the same as the carbon cost for energy generation more broadly. Using this metric, it is possible to achieve an effective ERE below 0.6 --- a significant energy savings overall.

This discussion generally assumes that the infrastructure costs (embodied carbon) for a heat exchange system is comparable to other waste heat disposal infrastructure, or that it is a relatively small component of the total embodied carbon cost of building infrastructure when a new data center is constructed or significant renovations are undertaken. Based on the components involved, this is likely a reasonable assumption in many areas of the world relevant to the WLCG.


%
\section{Sharing knowledge}
\label{sec:sitesharing}

One final source of computational resource waste that can be reduced to some degree is the rate of failing jobs.
Simply examining the failure rates across sites is insufficient to understand where issues lie, however.
It is critical in examining these failures to distinguish between ``site failures'' that are the responsibility of the site administrator or local configuration and ``software failures'' that are the responsibility of the production teams or users that have submitted the jobs.

Some sites offer unusual resources like very high memory nodes or GPUs, and therefore see different failure rates owing to different mixtures of workflows.
Similarly, some sites are configured to accept jobs at different rates: the Tier-0 computing site at CERN accepts mostly very well understood job configurations and therefore has a failure rate below 2\%, while general user analysis jobs lose about 15\% of the total wall clock time to failures.
Across the distributed production system, non-user jobs lose about 6\% of wall clock time to failures, an improvement from 10\% three to five years ago.

Worldwide, sites have many different hardware and software configurations that can affect operations.
The use of network-mounted storage or the size of a cache for software on the worker nodes, for example, can affect the running of jobs.
Comparing the rate of failures on different sites can therefore provide useful information to site administrators about what configuration changes might be made to the site to improve its efficiency.
To make this comparison fair, however, the variability of workloads across sites has to be accounted for.

\begin{figure}[tb]
\centering
\subfloat[]{
\includegraphics[width=0.9\textwidth]{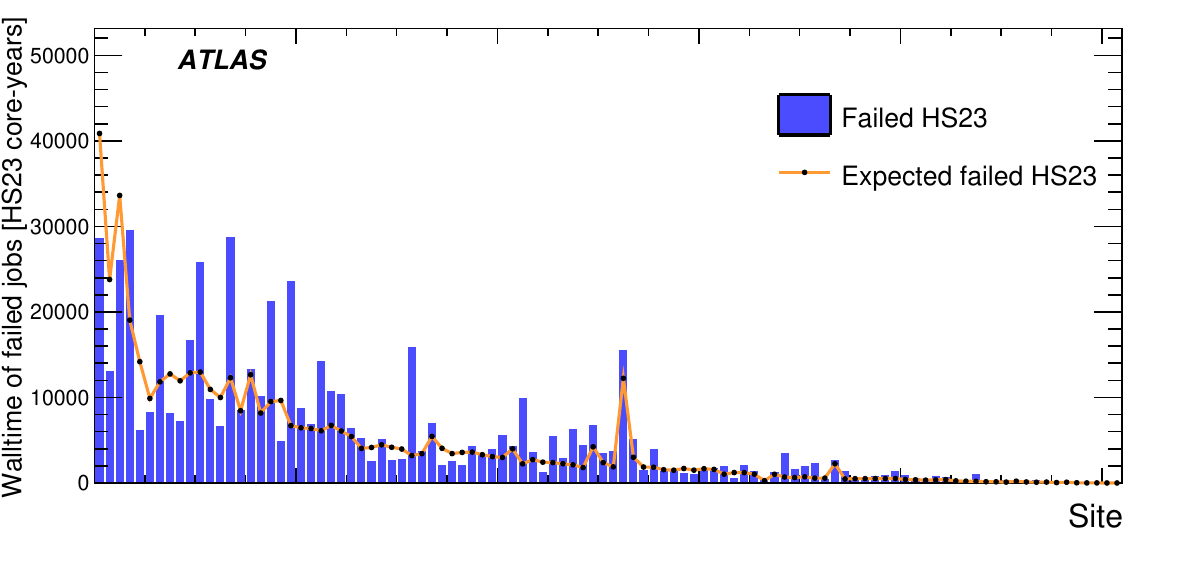}
} \\
\subfloat[]{
\includegraphics[width=0.9\textwidth]{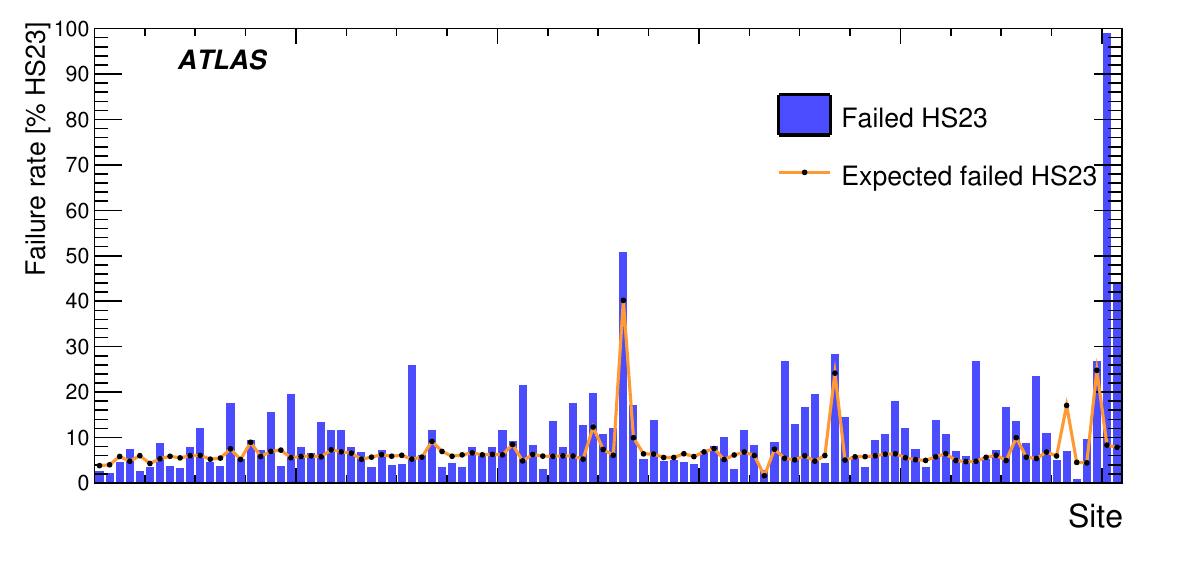}
}
\caption{(a) The computing time in HS23s at a given site lost to failing production (non-user) jobs in the last year. The observed loss is shown along with an expected loss based on the weighted world-average failure rates. (b) The same loss as a fraction of the HS23s provided by the site. In both cases, sites are ordered in the number of HS23 they contribute to ATLAS, from (left) most to (right) least.}
\label{fig:sitefailures}
\end{figure}

Figure~\ref{fig:sitefailures} shows a comparison of the wall clock time in HS23s at ATLAS sites lost to failing production jobs over the last year, both in absolute terms and as a fraction of the total compute time provided to ATLAS by the sites.
In addition to the observed lost wall clock time, an expected loss is shown, which is derived based on an average failure rate.
For each site, this expectation is calculated using the number of hours of each workload+core-count combination run at the site (e.g.~8-core data reconstruction or 64-core detector simulation) and applying a world-average failure rate for that workload+core-count combination.
If a site runs only well-understood workloads, the expected loss to failures should be very low.
If a site runs complex workloads known to be prone to failure, the expected loss to failures will be high.
If a site runs unique workload+core-count configurations, as might be the case for HPC systems that are capable of running jobs with very high core counts, then its expected failures will match its observed failures.
Because of their variability, user jobs are excluded from this comparison.

It is clear from this comparison that some sites have failures significantly below, and some significantly above, what would be expected based on their job composition.
This first study is sufficient to identify sites that are far from expectation.
The next step is to go in some detail through the site configurations for these sites to try to identify issues that might be leading to significantly higher- or lower-than-expected failures, and to produce a set of recommendations for the sites based on these comparisons.


%
\section{Summary and ongoing studies}
\label{sec:summary}

The studies described in this paper represent the current state of understanding of issues around sustainability and the environmental impact of computing in ATLAS.
Many more studies are underway.
This work is not in isolation: collaborations with a wide variety of groups are ongoing, including work within the wider WLCG, with the goal of broader recommendations in line with those from Section~\ref{sec:sites}, and with an eye towards a general life-cycle assessment for LHC computing.
As companies recognize the importance of these issues, embodied carbon estimates should become more reliable, allowing a more nuanced and complex model of carbon emissions and environmental impact.
Recognizing that many ongoing efforts and strategies in the experiment have an environmental impact is one of the key aspects of these examinations.
In preparation for the expansion of resources that will come over the next few years, the policies and practices of the experiment and the plans of the computing sites can be examined through the lens of environmental sustainability to provide a more complete picture of the impact of changes and new development.
The efforts in computing are only one aspect of the broader sustainability efforts within the ATLAS Collaboration, which are already proving profitable in other areas~\cite{MuonGasses}.

The computing landscape is also changing, and the environmental ``best path'' for the experiment is expected to evolve with some of those changes.
For example, as power grids worldwide decarbonize, the importance of embodied carbon (Scope 3) in comparison to operational carbon (Scope 2) will continue to rise.
Changes to hardware production may also change the calculation: chip makers worldwide are improving their sustainability practices as well as the efficiency of their hardware.
More significant changes may come as well, like changes to chip packaging to integrate more components inside a single package, which can improve power efficiency at the cost of modularity and hardware lifetimes.
Within the WLCG, further effort is needed to develop concrete scenarios for the evolution of these external factors, and to monitor their evolution in the coming years.

As the experiment diversifies its computing resources, it is important to consider carbon as one of the metrics by which new technologies are judged.
GPUs, for example, offer significant promise for the acceleration of workloads and for increasing the density of compute.
However, if they are not used with a sufficient load over their lifetime, the total carbon cost may be higher than the cost of simply buying additional CPUs.
The rise of machine learning and AI is driving significant increases in data center environmental impact worldwide, and several efforts are ongoing to understand and mitigate these impacts (see for example Refs.~\cite{AIEnergy,ImranAIPower,AIInferenceFuture}).
If commercial cloud providers are willing to offer more detailed power monitoring information, it might be possible to thoroughly evaluate new hardware technologies in power and carbon footprint terms without having to purchase and deploy them in a WLCG site.
In terms of site diversification, ATLAS already runs software on ``traditional'' high-throughput compute farms, HPC centers, commercial clouds, and personal computing resources through volunteer programs.
The carbon intensity of these various types of site should be evaluated where possible, in as complete a way as possible.
For example, HPC centers offer dense, high-performance computing, but on machines that often have short lifetimes.

The environmental impact of computing in ATLAS is expected to feature more prominently as the upgrades for the HL-LHC approach.
The large-scale expansion of resources offers an opportunity for major changes in purchasing patterns, usage, and policies.
These practices also serve as a testbed for future experiments, where computing challenges may be even more complex.


%

%
%
%
%
%
\section*{Acknowledgements}

%
%
%
%
%
%
%
%

%

%
%

%
%

We thank CERN for the very successful operation of the LHC and its injectors, as well as the support staff at
CERN and at our institutions worldwide without whom ATLAS could not be operated efficiently.

The crucial computing support from all WLCG partners is acknowledged gratefully, in particular from CERN, the ATLAS Tier-1 facilities at TRIUMF/SFU (Canada), NDGF (Denmark, Norway, Sweden), CC-IN2P3 (France), KIT/GridKA (Germany), INFN-CNAF (Italy), NL-T1 (Netherlands), PIC (Spain), RAL (UK) and BNL (USA), the Tier-2 facilities worldwide and large non-WLCG resource providers. Major contributors of computing resources are listed in Ref.~\cite{ATL-SOFT-PUB-2025-001}.

We gratefully acknowledge the support of ANPCyT, Argentina; YerPhI, Armenia; ARC, Australia; BMWFW and FWF, Austria; ANAS, Azerbaijan; CNPq and FAPESP, Brazil; NSERC, NRC and CFI, Canada; CERN; ANID, Chile; CAS, MOST and NSFC, China; Minciencias, Colombia; MEYS CR, Czech Republic; DNRF and DNSRC, Denmark; IN2P3-CNRS and CEA-DRF/IRFU, France; SRNSFG, Georgia; BMBF, HGF and MPG, Germany; GSRI, Greece; RGC and Hong Kong SAR, China; ICHEP and Academy of Sciences and Humanities, Israel; INFN, Italy; MEXT and JSPS, Japan; CNRST, Morocco; NWO, Netherlands; RCN, Norway; MNiSW, Poland; FCT, Portugal; MNE/IFA, Romania; MSTDI, Serbia; MSSR, Slovakia; ARIS and MVZI, Slovenia; DSI/NRF, South Africa; MICIU/AEI, Spain; SRC and Wallenberg Foundation, Sweden; SERI, SNSF and Cantons of Bern and Geneva, Switzerland; NSTC, Taipei; TENMAK, T\"urkiye; STFC/UKRI, United Kingdom; DOE and NSF, United States of America.

Individual groups and members have received support from BCKDF, CANARIE, CRC and DRAC, Canada; CERN-CZ, FORTE and PRIMUS, Czech Republic; COST, ERC, ERDF, Horizon 2020, ICSC-NextGenerationEU and Marie Sk{\l}odowska-Curie Actions, European Union; Investissements d'Avenir Labex, Investissements d'Avenir Idex and ANR, France; DFG and AvH Foundation, Germany; Herakleitos, Thales and Aristeia programmes co-financed by EU-ESF and the Greek NSRF, Greece; BSF-NSF and MINERVA, Israel; NCN and NAWA, Poland; La Caixa Banking Foundation, CERCA Programme Generalitat de Catalunya and PROMETEO and GenT Programmes Generalitat Valenciana, Spain; G\"{o}ran Gustafssons Stiftelse, Sweden; The Royal Society and Leverhulme Trust, United Kingdom.

In addition, individual members wish to acknowledge support from CERN: European Organization for Nuclear Research (CERN DOCT); Chile: Agencia Nacional de Investigaci\'on y Desarrollo (FONDECYT 1230812, FONDECYT 1230987, FONDECYT 1240864); China: Chinese Ministry of Science and Technology (MOST-2023YFA1605700, MOST-2023YFA1609300), National Natural Science Foundation of China (NSFC - 12175119, NSFC 12275265); Czech Republic: Czech Science Foundation (GACR - 24-11373S), Ministry of Education Youth and Sports (ERC-CZ-LL2327, FORTE CZ.02.01.01/00/22\_008/0004632), PRIMUS Research Programme (PRIMUS/21/SCI/017); EU: H2020 European Research Council (ERC - 101002463); European Union: European Research Council (ERC - 948254, ERC 101089007, ERC, BARD, 101116429), European Regional Development Fund (SMASH COFUND 101081355, SLO ERDF), Horizon 2020 Framework Programme (MUCCA - CHIST-ERA-19-XAI-00), European Union, Future Artificial Intelligence Research (FAIR-NextGenerationEU PE00000013), Italian Center for High Performance Computing, Big Data and Quantum Computing (ICSC, NextGenerationEU); France: Agence Nationale de la Recherche (ANR-21-CE31-0022, ANR-22-EDIR-0002); Germany: Baden-Württemberg Stiftung (BW Stiftung-Postdoc Eliteprogramme), Deutsche Forschungsgemeinschaft (DFG - 469666862, DFG - CR 312/5-2); China: Research Grants Council (GRF); Italy: Istituto Nazionale di Fisica Nucleare (ICSC, NextGenerationEU), Ministero dell'Università e della Ricerca (NextGenEU PRIN20223N7F8K M4C2.1.1); Japan: Japan Society for the Promotion of Science (JSPS KAKENHI JP22H01227, JSPS KAKENHI JP22H04944, JSPS KAKENHI JP22KK0227, JSPS KAKENHI JP25H01291, JSPS KAKENHI JP25K01023); Norway: Research Council of Norway (RCN-314472); Poland: Ministry of Science and Higher Education (IDUB AGH, POB8, D4 no 9722), Polish National Science Centre (NCN 2021/42/E/ST2/00350, NCN OPUS 2023/51/B/ST2/02507, NCN OPUS nr 2022/47/B/ST2/03059, NCN UMO-2019/34/E/ST2/00393, UMO-2022/47/O/ST2/00148, UMO-2023/49/B/ST2/04085, UMO-2023/51/B/ST2/00920, UMO-2024/53/N/ST2/00869); Portugal: Foundation for Science and Technology (FCT); Spain: Ministry of Science and Innovation (MCIN \& NextGenEU PCI2022-135018-2, MICIN \& FEDER PID2021-125273NB, RYC2019-028510-I, RYC2020-030254-I, RYC2021-031273-I, RYC2022-038164-I); Sweden: Carl Trygger Foundation (Carl Trygger Foundation CTS 22:2312), Swedish Research Council (Swedish Research Council 2023-04654, VR 2021-03651, VR 2022-03845, VR 2022-04683, VR 2023-03403, VR 2024-05451), Knut and Alice Wallenberg Foundation (KAW 2018.0458, KAW 2022.0358, KAW 2023.0366); Switzerland: Swiss National Science Foundation (SNSF - PCEFP2\_194658); United Kingdom: EPSRC (CANDISE, EP/X038858/1), Leverhulme Trust (Leverhulme Trust RPG-2020-004), Royal Society (NIF-R1-231091); United States of America: U.S. Department of Energy (ECA DE-AC02-76SF00515), NSF (CNS-1901466), Neubauer Family Foundation; VMware University Research Fund.

%
%


%
%
%
\printbibliography
\clearpage
 
\begin{flushleft}
\hypersetup{urlcolor=black}
{\Large The ATLAS Collaboration}

\bigskip

\AtlasOrcid[0000-0002-6665-4934]{G.~Aad}$^\textrm{\scriptsize 104}$,
\AtlasOrcid[0000-0001-7616-1554]{E.~Aakvaag}$^\textrm{\scriptsize 17}$,
\AtlasOrcid[0000-0002-5888-2734]{B.~Abbott}$^\textrm{\scriptsize 123}$,
\AtlasOrcid[0000-0002-0287-5869]{S.~Abdelhameed}$^\textrm{\scriptsize 119a}$,
\AtlasOrcid[0000-0002-1002-1652]{K.~Abeling}$^\textrm{\scriptsize 55}$,
\AtlasOrcid[0000-0001-5763-2760]{N.J.~Abicht}$^\textrm{\scriptsize 49}$,
\AtlasOrcid[0000-0002-8496-9294]{S.H.~Abidi}$^\textrm{\scriptsize 30}$,
\AtlasOrcid[0009-0003-6578-220X]{M.~Aboelela}$^\textrm{\scriptsize 45}$,
\AtlasOrcid[0000-0002-9987-2292]{A.~Aboulhorma}$^\textrm{\scriptsize 36e}$,
\AtlasOrcid[0000-0001-5329-6640]{H.~Abramowicz}$^\textrm{\scriptsize 157}$
% ...
\bigskip
\\

% ...
$^{17}$Department for Physics and Technology, University of Bergen, Bergen; Norway.\\
% ...
$^{30}$Physics Department, Brookhaven National Laboratory, Upton NY; United States of America.\\
% ...
$^{36}$$^{(a)}$Facult\'e des Sciences Ain Chock, Universit\'e Hassan II de Casablanca;$^{(b)}$Facult\'{e} des Sciences, Universit\'{e} Ibn-Tofail, K\'{e}nitra;$^{(c)}$Facult\'e des Sciences Semlalia, Universit\'e Cadi Ayyad, LPHEA-Marrakech;$^{(d)}$LPMR, Facult\'e des Sciences, Universit\'e Mohamed Premier, Oujda;$^{(e)}$Facult\'e des sciences, Universit\'e Mohammed V, Rabat;$^{(f)}$Institute of Applied Physics, Mohammed VI Polytechnic University, Ben Guerir; Morocco.\\
% ...
$^{45}$Physics Department, Southern Methodist University, Dallas TX; United States of America.\\
% ...
$^{49}$Fakult\"{a}t Physik , Technische Universit{\"a}t Dortmund, Dortmund; Germany.\\
% ...
$^{55}$II. Physikalisches Institut, Georg-August-Universit\"{a}t G\"ottingen, G\"ottingen; Germany.\\
% ...
$^{104}$CPPM, Aix-Marseille Universit\'e, CNRS/IN2P3, Marseille; France.\\
% ...
$^{119}$$^{(a)}$New York University Abu Dhabi, Abu Dhabi;$^{(b)}$United Arab Emirates University, Al Ain; United Arab Emirates.\\
% ...
$^{123}$Homer L. Dodge Department of Physics and Astronomy, University of Oklahoma, Norman OK; United States of America.\\
% ...
$^{157}$Raymond and Beverly Sackler School of Physics and Astronomy, Tel Aviv University, Tel Aviv; Israel.\\
% ...
$^{a}$ Also at Affiliated with an institute formerly covered by a cooperation agreement with CERN.\\
$^{b}$ Also at An-Najah National University, Nablus; Palestine.\\
$^{c}$ Also at Borough of Manhattan Community College, City University of New York, New York NY; United States of America.\\
$^{d}$ Also at Center for Interdisciplinary Research and Innovation (CIRI-AUTH), Thessaloniki; Greece.\\
$^{e}$ Also at Centre of Physics of the Universities of Minho and Porto (CF-UM-UP); Portugal.\\
$^{f}$ Also at CERN, Geneva; Switzerland.\\
% ...

\end{flushleft}


%

%
%
%
%

%
%
%
%

\end{document}